\documentclass[citeautoscript,floatfix,aps,prx,twocolumn,superscriptaddress
]{revtex4-2}

\usepackage{gensymb}
\usepackage{graphicx}
\usepackage{dcolumn}
\usepackage{bm}
\usepackage{float}
\usepackage{amsmath}

\newcommand{\AG}[1]{\textcolor{red}{\textbf{[AG]} #1}}

\newcommand{\CR}[1]{\textcolor{blue}{\textbf{[CR]} #1}}

\usepackage[breaklinks=true,colorlinks,citecolor=blue,linkcolor=blue,urlcolor=blue]{hyperref}
\usepackage{comment}
\usepackage{braket}

\usepackage[
commandnameprefix=always]{changes}
\definechangesauthor[name=Antoine, color=purple]{AG}

\begin{document}
\title{Superconductivity in the two-dimensional Hubbard model\\ revealed by neural quantum states}

\author{Christopher Roth}
\affiliation{Center for Computational Quantum Physics, Flatiron Institute, New York 10010, USA}

\author{Ao Chen}
\affiliation{Center for Computational Quantum Physics, Flatiron Institute, New York 10010, USA}
\affiliation{Division of Chemistry and Chemical Engineering, California Institute of Technology, Pasadena, California 91125, USA}
\affiliation{Center for Electronic Correlations and Magnetism, University of Augsburg, 86135 Augsburg, Germany}

\author{Anirvan Sengupta}
\affiliation{Center for Computational Quantum Physics, Flatiron Institute, New York 10010, USA}
\affiliation{Center for Computational Mathematics, Flatiron Institute, 162 5th Avenue, New York, New York 10010, USA}
\affiliation{Department of Physics and Astronomy, Rutgers University, Piscataway, New Jersey 08854, USA}

\author{Antoine Georges}
\affiliation{Center for Computational Quantum Physics, Flatiron Institute, New York 10010, USA}
\affiliation{Coll{\`e}ge de France, 11 place Marcelin Berthelot, 75005 Paris, France}
\affiliation{CPHT, CNRS, {\'E}cole Polytechnique, IP Paris, F-91128 Palaiseau, France}
\affiliation{DQMP, Universit{\'e} de Gen{\`e}ve, 24 quai Ernest Ansermet, CH-1211 Gen{\`e}ve, Suisse}

\date{\today}

\begin{abstract}
Whether the ground state of the square lattice Hubbard model exhibits superconductivity remains a major open question, 
central to understanding high temperature cuprate superconductors and
ultra-cold fermions in optical lattices.
Numerical studies have found evidence for stripe-ordered states and superconductivity at strong coupling
but the phase diagram remains controversial.
Here, we show that one can resolve the subtle energetics of metallic, superconducting, 
and stripe phases using a new class 
of neural quantum state (NQS) wavefunctions 
that extend hidden fermion determinant states to Pfaffians. We simulate several hundred electrons using fast Pfaffian algorithms 
allowing us to measure off-diagonal long range order.
At strong coupling and low hole-doping, we find that a non-superconducting filled stripe phase prevails, while superconductivity coexisting with partially-filled stripes is stabilized by a negative next neighbor hopping 
$t^\prime$, with $|t^\prime|>0.1$. 
At larger doping levels, we introduce momentum-space correlation functions to mitigate finite size effects that arise from weakly-bound pairs. These provide evidence for uniform d-wave superconductivity at $U=4$, even when $t'=0$. 
%
Our results highlight the potential of NQS approaches, and provide a fresh perspective on superconductivity in the square lattice Hubbard model.  
\end{abstract}

\maketitle

\begin{figure*}[t] 
    \centering
    \includegraphics[width=0.9\linewidth]{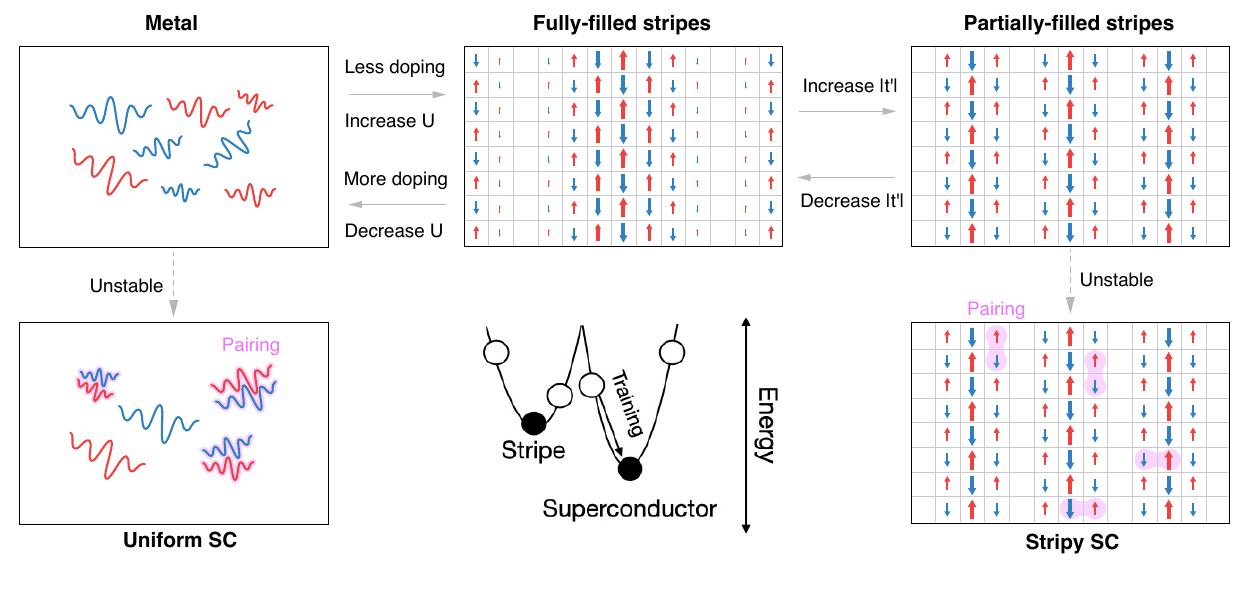}
    \caption{Depiction of the phases described in this work. The top row shows non-superconducting ``normal state" phases. The colors red/blue represent spin up and down. The wavy lines represent wave-like electrons that occupy momentum eigenstates, while the up/down symbols represent spatially localized electrons. 
    The horizontal arrows specify the changes in Hamiltonian couplings that move the ground state between different phases. For the metal and partially-filled stripe, we find that a small proportion of the electrons form itinerant singlet pairs, giving rise to superconductivity.  This instability towards pairing is shown by the downward pointing vertical arrows.  In the middle we show a cartoon of our optimization strategy. Each optimization is run with several different mean-field initial conditions (empty circles). These optimizations converge to distinct phases (black circles) that are not sensitive to the details of the initialization. 
    \label{fig: cartoon main}
    }
\end{figure*}


\section{Introduction}


It was suggested early on~\cite{Anderson_Science87_RVB, KotliarSuperexchange, DopingMottInsulator, Ruckenstein_mean_field_J} that antiferromagnetic superexchange in the presence 
of strong repulsive interactions may be responsible for the high-temperature superconductivity 
(SC) observed in copper oxide compounds (cuprates). 
In this context, the two-dimensional square lattice Hubbard model provides the simplest framework in which this mechanism 
can be put to the test~\cite{Qin_ARCMP22_Hubbard,arovas_ARCMP_2022}.


Establishing the ground-state phase diagram of this model has proven very challenging because of the presence of phenomenologically distinct states with extremely close energies. 
In recent years, computational methods have nonetheless made great strides on this problem, so that some properties are now established without a shred of doubt. 
For example, all computational methods agree that at strong coupling and 
$1/8$-doping, the ground-state of the Hubbard model with only nearest-neighbor hopping hosts a charge and spin-density wave - with doped holes forming insulating stripes~\cite{zheng2017stripe,xu2022stripe,Qin_PRX20_Hubbard,Xu_Science24_Hubbard,gullmillis_natphys_2015,peters_2014}. 
All methods also agree that the model displays a strong response to an applied 
d-wave pairing field in a range of doping levels~\cite{kancharla_2008,gullmillis_natphys_2015,Qin_PRX20_Hubbard}. 
More broadly, there is now a fairly good agreement on the possible competing or 
intertwined long-range orders.


We are now at an exciting frontier where the goal is to assess which of these possible 
phases truly has the lowest energy, and computational methods 
are becoming accurate enough to address this challenge. 
An outstanding question, for example, is whether the insulating stripe ground-state 
can give way to a d-wave superconductor when a next nearest-neighbor hopping $t^\prime$ is turned on. 
A remarkable recent work ~\cite{Xu_Science24_Hubbard}
combining constrained-path auxiliary field quantum Monte Carlo (CP-AFQMC) and density-matrix renormalization group (DMRG)  simulations with careful 
consideration of boundary conditions 
recently answered this question affirmatively, yet this still remains a source of debate
~\cite{Jiang4leg,Jiang_Science19_DMRG-Hubbard,zhang_vondelft_2025,jiang_6legs_2024, DQMC_Chen, VMC_Hubbard_tprime_Lechiara}. 


Another thread of research is understanding the pure Hubbard model, with $t^\prime = 0$, 
for which cold fermionic gases in optical lattices provide an analog quantum simulation platform. 
These experiments have recently reached unprecedentedly low temperatures where 
collective effects are expected to emerge\cite{xu_nature_2025,chalopin_2024,kendrick_2025}. 
While it is known that this model has a superconducting phase with an exponentially small 
order parameter in the $U\ll t$ limit~\cite{Raghu2010weakcoupling, KohnLuttinger,Deng_epl_2015} 
it is important to establish whether pairing becomes significantly larger when on-site repulsion is increased. Answering this question, which is still an ongoing source of debate~\cite{Qin_PRX20_Hubbard,imadaSCpurehubbard,sorella2023systematically}, may help to inform future choices for experiments.


We examine both of these regimes, the $t-t^\prime$ model at strong coupling, and the pure Hubbard model in a range of coupling favorable for d-wave superconductivity. Our simulations use the recently introduced hidden fermion Pfaffian states (HFPS) \cite{chen2025pfaffian}. 
Pfaffian wavefunctions provide an expressive ansatz which can be naturally combined 
with deep neural  networks in order to capture charge, spin and 
superconducting orders and their interplay.
Our work, in conjunction with reference \cite{Gu_arxiv25_NQS-Hubbard}, extends fermionic neural quantum states (fNQS) simulations to much larger systems of several hundred electrons.



First, we simulate the $t-t^\prime$ model at $U/t=8$ and $1/8$-doping. In agreement with other numerical methods, we find an insulating stripe phase at $t^\prime =0$~\cite{xu2022stripe,simons_colab_stripes_zheng}, which transitions to a shorter period ("partially-filled"~\footnote{The filling of a stripe refers to the ratio between the number of holes per domain wall and the number of sites per domain wall}) stripe when
a large enough negative next neighbor hopping is turned on \cite{Xu_Science24_Hubbard, Gu_arxiv25_NQS-Hubbard,  huang2018stripe}. Notably, we find that this partially-filled stripe exhibits superconductivity with a pairing order parameter of $O(10^{-2})$, in sharp agreement with the recent handshake between DMRG and CP-AFQMC \cite{Xu_Science24_Hubbard}. In contrast with the handshake study, which measures the linear response to an applied pairing field, we look for superconductivity by measuring the long-range behavior of pair correlation functions. 
Therefore, this indicates an exciting convergence between studies which differ in both their computational schemes and methods of detecting pairing.

Next, we investigate whether superconductivity with a sizable order parameter can exist without $t^\prime$. Here we focus on intermediate coupling $U/t=4$ at $1/6$-doping where most methods agree that stripe ordering does not take place \cite{xu2022stripe, sorella2023systematically}.  
We present evidence that the ground state has fairly strong d-wave superconductivity for these parameters, which may be detectable in cold atom experiments. 
By running accurate simulations on large, symmetric geometries, we show that unlike the $t^\prime \ne 0$ case, this superconductor hosts pairs with a very long coherence length. 
This leads to stronger finite-size effects than in the strong coupling regime, which motivates us to introduce momentum-space correlation functions that are better suited to the investigation 
of regimes with weakly bound pairs. 
This observation resolves many of the disagreements among different computational studies, as the local 
d-wave order parameter only captures a small fraction of the total pairing for weakly-bound pairs. Therefore, although we find robust superconductivity, it is not incompatible with the previous finding that the local d-wave order parameter is quite small~\cite{Qin_PRX20_Hubbard}.

Using a flexible Pfaffian-based ansatz and low-rank update algorithms, we accurately simulate the square lattice Hubbard model on large symmetric geometries, carefully accounting for finite size effects. We provide new insight into the intertwined orders and the nature of pairing, heightening our understanding of how high-$T_c$ superconductivity may emerge from microscopic interactions.


\section{Methods}

\subsection{Hubbard model}
We simulate the square lattice Hubbard model with an additional next-neighbor interaction, 
\begin{equation}
    {\hat H}_{\mathrm{Hub}} = U \sum_{\bf x} {\hat c}^\dagger_{{\bf x},\uparrow} {\hat c}_{{\bf x},\uparrow} {\hat c}^\dagger_{{\bf x}, \downarrow} {\hat c}_{{\bf x},\downarrow} - \sum_{{\bf x}, {\bf x'}}\sum_\sigma t_{{\bf x} {\bf x'}} {\hat c}^\dagger_{{\bf x},\sigma} {\hat c}_{{\bf x'},\sigma}
\end{equation}

Here $U$ represents an energy penalty from Coulomb repulsion when two electrons share a site, 
and the hopping amplitudes $t_{{\bf x}, {\bf x'}}$ are associated with the kinetic energy gain, with 
$t_{{\bf x}, {\bf x'}}=t$ for nearest-neighbor hopping and 
$t_{{\bf x}, {\bf x'}}=t^\prime$ for next nearest-neighbor hopping. 
Without loss of generality, we set $t=1$. 

\subsection{Hidden Fermion Pfaffian States} \label{sec: HFPS}

We use the recently introduced hidden fermion Pfaffian states (HFPS) \cite{chen2025pfaffian} as a variational wavefunction in order to simulate different regimes of the Hubbard model to high accuracy. This method extends the hidden fermion method \cite{Moreno_PNAS22_HFDS} to wavefunctions with pairing. 

The HFPS ansatz can be written in the following form, 
\begin{equation} \label{eqn: HFPS unprojected}
\psi({\bf n}) = J({\bf n})\, \textrm{Pf} \Bigg({\bf n} \star \Big(F + H({\bf n}) A H^T({\bf n}) \Big)  \star {\bf n} \Bigg),     
\end{equation}
where $J({\bf n})$ is a neural network Jastrow, F is the mean field orbital matrix, and $H({\bf n}) A H^T({\bf n})$ can be viewed as a backflow correction. The matrix $A$, which is an antisymmetric matrix that describes the coupling between hidden fermions, $H({\bf n})$, has shape $2{\tilde n} \times 2{\tilde n}$, where ${\tilde n}$ is the number of hidden fermions. If the number of hidden fermions is much smaller than the number of electrons, ${\tilde n} << n$, a low rank update scheme can be used to compute the ratio between the wavefunction amplitudes of Fock states that differ by a small number of electron hops. This massively speeds up Monte Carlo and allows us to access large system sizes. Both $J({\bf n})$ and $H({\bf n})$ are parametrized by a single group-convolutional neural network \cite{Cohen_arxiv16_GCNN, Roth_PRB23_GCNN}. The first ${\tilde n}$ feature vectors are used to compute $H({\bf n})$, while the remaining are symmetry averaged to compute $J({\bf n})$. Finally, the operation $\star$ selects the orbitals that are occupied by fermions in the Fock state (often referred to as `slicing') \cite{Moreno_PNAS22_HFDS,chen2025pfaffian}.

We can control the symmetry breaking unit cell of our wavefunction by choosing an appropriate block-circulant structure for $F$. This allows us to target particular density wave configurations, such as filled and half-filled stripes. 

In order to improve the final variational accuracy, $\psi({\bf n})$ is symmetry projected, so that the wavefunction has the full symmetry of the Hamiltonian,
\begin{equation}
    \langle {\bf n} | \psi \rangle = \sum_{g \in G} \psi({g {\bf n}}).
\end{equation}
Here, $G$ represents the group of symmetry operations that are broken by the unit cell. See the reference \cite{Moreno_PNAS22_HFDS} for details about the hidden fermion formalism and \cite{chen2025pfaffian} about how it is generalized to Pfaffians. 

\subsection{Expressivity of the Pfaffian} \label{sec: appendix Pfaffian expressivity}


It has been well established that the Hubbard model has many competing phases of matter that are extremely close in energy \cite{xu2022stripe}. These phases include magnetically ordered insulators, metals and potentially superconductors. There is also emerging evidence that more exotic phases, which can have both long ranged density wave and pairing order, can emerge in the presence of a next neighbor hopping $t^\prime$ \cite{Xu_Science24_Hubbard}. We emphasize that the Pfaffian,
\begin{equation} 
    \langle {\bf n} | \psi_{pf} \rangle = \textrm{Pf} \Big({\bf n} \star  F \star {\bf n} \Big),   
\end{equation}
can represent all of these phases at the mean field level. 


The Pfaffian is a solution to the Bogoliubov–de Gennes (BdG) Hamiltonian which can be written in the following form,
\begin{equation*} \label{eqn: BDG}
H = \frac{1}{2} \sum_{{\bf x}, {\bf x^\prime}} \Psi_{\bf x}^\dagger \begin{pmatrix} h^{\sigma, \sigma^\prime}_{{\bf x}, {\bf x^\prime}} & \Delta^{\sigma, \sigma^\prime}_{{\bf x}, {\bf x^\prime}}  \\ -(\Delta^{\sigma, \sigma^\prime}_{{\bf x}, {\bf x^\prime}})^* & -(h_{{\bf x}, {\bf x^\prime}}^{\sigma, \sigma^\prime})^T 
\end{pmatrix}
\Psi_{\bf x'}
\end{equation*}
where $h^{\sigma, \sigma^\prime}_{{\bf x}, {\bf x^\prime}}$ and $\Delta^{\sigma, \sigma^\prime}_{{\bf x}, {\bf x^\prime}}$ are $2 \times 2$ matrices which describe hopping and pairing respectively between up and down spins, and  $\Psi_{\bf x} = (c_{\bf x \uparrow}, c_{\bf x \downarrow}, c^\dagger_{\bf x \uparrow}, c^\dagger_{\bf x \downarrow} )$ is a Nambu spinor. The general solution can have spin and density correlations $\langle c^\dagger_{\bf x, \sigma} c_{\bf x, \sigma'} \rangle \ne 0$,  as well as singlet $\langle c_{\bf x, \uparrow} c_{\bf x^\prime, \downarrow} \rangle \ne 0$ and triplet $\langle c_{\bf x, \uparrow} c_{\bf x^\prime, \uparrow} \rangle \ne 0$ pairing. 

Therefore, the Pfaffian can represent phases with intertwined charge, spin, superconducting orders, making it an ideal starting point for flexibly capturing the phases of the square lattice Hubbard model. 

The Slater determinant represents a subclass of solutions to the BdG Hamiltonian where $\Delta_{\bf x, \bf x'}^{\sigma, \sigma^\prime} = 0$. While it cannot represent superconducting order directly, in some cases unitary particle-hole transformations can be used to achieve this goal. However, our contention is that Pfaffian provides a more natural and expressive framework.

\subsection{Optimization strategy}

It is well established that NQS wavefunctions have different local minima of the variational energy, corresponding to different phases of a given Hamiltonian \cite{Astrakhantsev_PRX21_pyrochlore, smith2024unified}. 
In order to ensure that we find the phase that is a global minimum, we run several controlled optimizations starting from distinct mean field solutions. We run each trajectory by initializing $F$ with a plausible 
ground-state order, and setting $J({\bf n}) = 1$, $A=0$ so that our initial wavefunction is the 
mean-field Pfaffian in Eqn. \ref{eqn: HFPS unprojected}. The mean field orbitals, $F$, are trained with exact gradient on the following Hamiltonian, 
\begin{equation} \label{eqn: mean field ham}
    \hat{\mathcal{H}}_{0} = \hat{\mathcal{H}}_{\mathrm{Hub}}(U_0) + \Delta_{0} \sum_{\boldsymbol{\delta}} \textrm{sign}({\boldsymbol{\delta}}) (c^\dagger_{{\bf x} + \boldsymbol{\delta} \uparrow} c^\dagger_{\bf x \downarrow}  + h.c.),
\end{equation}
where $\boldsymbol{\delta}$ runs over the four nearest neighbor displacements and $\textrm{sign}(\boldsymbol{\delta})$ is $+1$ for vertical bonds and $-1$ for horizontal bonds. The d-wave pinning field is necessary to initialize the wavefunction with pairing, as the best mean-field solution to $\mathcal{H}_{\mathrm{Hub}}$ always has $\langle c_{\bf x, \sigma} c_{\bf x \sigma^\prime}\rangle = 0$. This is because the pairing correlations only enter quadratically through the onsite repulsion; therefore they are zero for $U_0 > 0$ at the mean-field level. 

We can find stripes of different periodicity by constraining $F$ to have particular block-circulant forms, while $U_0$ and $\Delta_0$ can be used to control the strength of the magnetism and superconductivity. We have experimented with initial conditions in both regimes and find that there are distinct local minima which are close in energy but have different long range order. However, once the optimization converges to a particular ordering, the final solution is fairly insensitive to the details of the initialization, as documented below. 

\subsection{Detecting long-range order from correlation functions}

Our simulations are performed on finite size lattices -- clusters of $N$ sites with periodic boundary conditions -- where spin, charge and lattice symmetries are preserved. Therefore, we must detect spontaneously broken symmetries by looking for long range order in two-body correlations. In this study we focus on spin,
\begin{equation} \label{eqn: spin_correlation}
    C_s(\mathbf{x}) = \frac{1}{N} \sum_{\bf x'} \Big[ \braket{\hat{S}^z_\mathbf{x + x'} \hat{S}^z_\mathbf{x'}}
    - \braket{\hat{S}^z_\mathbf{x + x'}} \braket{\hat{S}^z_\mathbf{x'}} \Big],
\end{equation}
charge,
\begin{equation} \label{eqn: charge_correlation}
    C_c(\mathbf{x}) = \frac{1}{N} \sum_{\bf x'} \Big[ \braket{\hat{n}^z_\mathbf{x + x'} \hat{n}_\mathbf{x'}}
    - \braket{\hat{n}_\mathbf{x + x'}} \braket{\hat{n}^z_\mathbf{x'}} \Big],
\end{equation}
and pair correlations
\begin{equation} \label{eqn: d-wave pair-correlation}
    C_p(\mathbf{x}) = \frac{1}{N} \sum_{\bf x'} \Big[ \braket{\hat d^\dagger(\mathbf{x + x'}) \hat d(\mathbf{x'})} - D_0({\bf x + x^\prime}, {\bf x^\prime}) \Big]
\end{equation}
for d-wave symmetrized pairs \footnote{An alternative definition of the pair correlation is given in references \cite{Qin_PRX20_Hubbard, Xu_Science24_Hubbard}, which defines a Cooper pair as $({\hat c}_{\bf x, \uparrow} {\hat c}_{\bf x', \downarrow} - {\hat c}_{\bf x, \downarrow} {\hat c}_{\bf x', \uparrow})/\sqrt{2}$. Relative to our definition, this gives a factor of 2 enhancement of the pair correlation function. }  
\begin{equation} \label{eqn: d-wave order param}
    \hat d(\mathbf{x}) = \frac{1}{4} \sum_{\boldsymbol{\delta}}\mathrm{sign}(\boldsymbol{\delta}) 
    \hat c_{\mathbf{x},\uparrow} \hat c_{\mathbf{x}+\boldsymbol{\delta}, \downarrow}.
\end{equation}
Here $\sum_{\boldsymbol{\delta}}$ runs over nearest neighbors with $\mathrm{sign}(\boldsymbol{\delta}) = +1$ for horizontal neighbors and $-1$ for vertical neighbors. We subtract $D_0({\bf x}, {\bf x'})$, the normal state contribution from uncorrelated hopping:  \footnote{The normal state contribution also contains a term that mixes spins $\langle c^\dagger_{\bf x, \uparrow} c_{\bf x, \downarrow} \rangle$ but this is zero for our simulations as they conserve $S^z$}
\begin{equation*}
    D_0({\bf x}, {\bf x^\prime}) = \frac{1}{16} \langle {\hat c}^\dagger_{\bf x, \uparrow} {\hat c}_{\bf x', \uparrow} \rangle \sum_{\boldsymbol \delta, \boldsymbol \delta^\prime} \textrm{sign}(\boldsymbol{\delta})  \textrm{sign}({\boldsymbol \delta}^\prime) \langle {\hat c}^\dagger_{\bf x + \boldsymbol \delta, \downarrow}  {\hat c}_{\bf x' + \boldsymbol \delta^\prime, \downarrow}  \rangle 
\end{equation*}
For the phases in Fig.~\ref{fig: cartoon main}, the metal has no long range order, and the stripe phases have long-range charge and spin order. This manifests as a periodic modulation of the charge and spin correlation functions. When there is superconductivity, we expect an additional long range d-wave pairing order. This means that the spatial integral of the pair correlation function is proportional to the 
size of the system at large sizes and hence that the average pair correlation defined as:
\begin{equation}
{\bar C}_p \equiv \frac{1}{N} \sum_{\bf x} C_p({\bf x})
\end{equation}
is non-zero in the thermodynamic limit:
\begin{equation}
    \lim_{N \rightarrow \infty} {\bar C}_p  > 0
\end{equation}

\begin{figure*}[t] 
    \centering
    \includegraphics[width=0.9\linewidth]{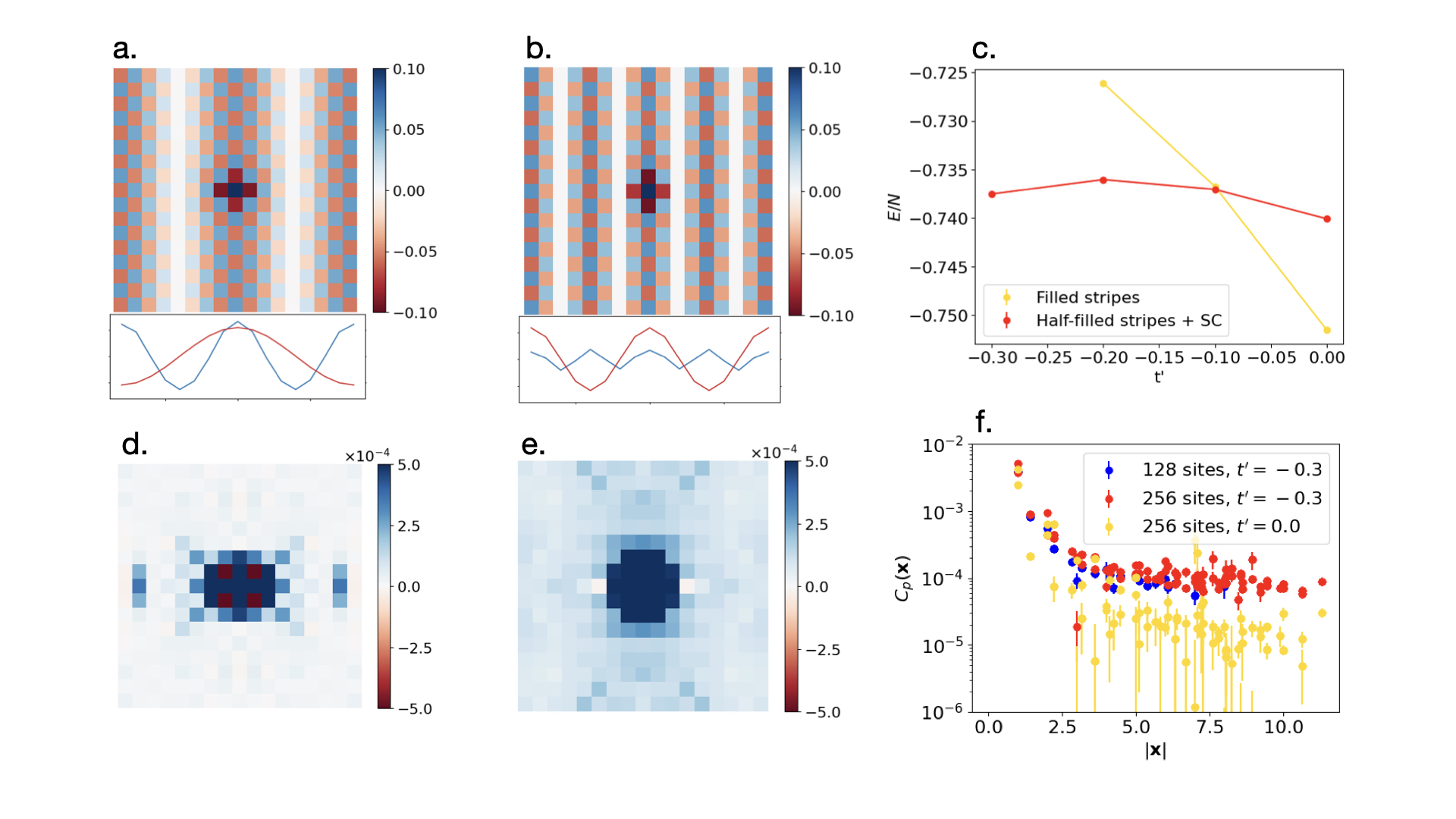}
    \caption{{\bf Stripes and stripy superconductors in the $t-t^\prime$ model at strong coupling}. Simulations are done on a $16 \times 16$ lattice at  $U=8$ and $\delta = 1/8$. All color plots are clipped at a maximum value shown in the color scale for ease of visualization. {\bf a.-\bf b.} Charge and spin correlation functions of the filled stripe at $t^\prime=0$  ({\bf a.}) and a half-filled stripe at $t^\prime = -0.3$ ({\bf b.}). The spin correlation function is plotted on top, while one-dimensional cuts of the charge, $C_c(x,y=8)$, and sublattice-transformed spin $(-1)^{x } C_s(x,y=8)$ correlations are plotted on the bottom. {\bf c.} Energies of the filled stripe and half-filled stripe on a $256$ site cluster as a function of  $t^\prime$. The statistical error bars are smaller than the data points. {\bf d.-\bf e.} Color plot of  pair correlations, $C_p({\bf x})$ at $t^\prime = 0$ ({\bf d.}) and $t^\prime = -0.3$ ({\bf e.}) {\bf f.} Pair correlations as a function of scalar displacement $|{\bf x}|$ for $t^\prime = 0$ on a $256$ site cluster (gold), and for $t^\prime = -0.3$ on $128$ (blue) and $256$ (red) site clusters.
    \label{fig: stripemain}
    }
\end{figure*}

\section{Phase transition between filled stripes and partially-filled stripes with superconductivity 
\label{sec: stripy SC}
}

First we study the Hubbard model at $U=8$ with hole-doping $\delta = 1/8$, considering the phase diagram as a next neighbor hopping, $t^\prime < 0$, is turned on. This is a relevant model for hole-doped cuprates 
such as La$_{2-x}$Sr$_x$CuO$_4$ or YBa$_2$Cu$_3$O$_{7-x}$ which typically have significant negative next-neighbor hopping, $-0.4 < t^\prime < -0.1$ \cite{pavarini2001band}. 
Building on prior work \cite{Gu_arxiv25_NQS-Hubbard, huang2018stripe, xu2022stripe} 
, we consider two competing phases, a filled-stripe with a charge (spin) modulation of period $\lambda_{c(s)} = 1(2)/\delta$, and a half-filled stripe with a charge (spin) modulation of period $\lambda_{c(s)} = \frac{1}{2} \times 1(2)/\delta$. 

Since these phases break different symmetries, they are likely separated by a first order phase transition. Therefore, once we have a converged model in each phase, we use transfer learning to compare the energies as $t^\prime$ is varied. Using a large $256$ site simulation cluster, we find a transition from a filled stripe to a half-filled stripe at $t^\prime = -0.1$ 
(Fig.\ref{fig: stripemain}c).  

The details of the stripe order after training are shown in Fig.\ref{fig: stripemain}a.-b. for $t^\prime = 0$ and $t^\prime = -0.3$ respectively. At $t^\prime=0$, we see that the spin correlations have period $16$ and the charge correlations have period 8, corresponding to a filled stripe. On the other hand, at  $t^\prime=-0.3$ the charge and spin periods are half as long, and therefore the domain walls are only half-filled with holes. While the amplitude of the spin modulation is roughly the same between the phases, the half-filled stripe has weaker charge modulation. 

In figure \ref{fig: stripemain}d.-f. we show the d-wave pair correlations for both $t^\prime =0$ and $t^\prime = -0.3$. We see in \ref{fig: stripemain}f. that the pair correlations for $t^\prime = -0.3$ plateau to a value of $10^{-4}$, in line with an order parameter of  $\mathcal{O}(10^{-2})$. This value is qualitatively consistent with the results of the recent CP-AFQMC+DMRG handshake collaboration \cite{Xu_Science24_Hubbard}, which also finds superconductivity coexisting with partially-filled stripes. We compare the pair correlations on a $128$ and a $256$ site cluster, finding that the asymptotic value is fairly consistent. This indicates that the $256$ site cluster is likely large enough to resolve off-diagonal long range order. In contrast, the pair correlations of the filled stripe phase decay to at least an order-of-magnitude smaller value and are nearly zero within our statistical error. Therefore, we expect that there is very little or no superconductivity in this phase (which is likely an insulator).

In order to show that the finding of superconductivity in the half-filled stripe is not biased by initialization, we document how the pair correlations in the half-filled stripe change with the strength of the d-wave pairing field $\Delta_0$ used to initialize the orbitals. 


\begin{table}[H]
    \centering
    \begin{tabular}{|c|c|c|c|}
    \hline 
          $\Delta_0$ &0.0& 0.2& 0.5 \\ \hline
          E/N &  -0.72452(5) &  -0.73338(4)& \textbf{-0.73752(4)}\\ \hline
          ${\bar C}_p(|{\bf x}| > 3)$&$-2(2) \times 10^{-6}$& $1.35(2)  \times 10^{-4}$& $\bf{1.06(2) \times 10^{-4}}$\\
     \hline     
     \end{tabular}
    \caption{Energies and average pair correlations (for pair separations of a distance larger than 3) for the partially-filled stripe at $U=8,\delta=1/8$ and $t^\prime = -0.3$  initialized with different pinning fields}
    \label{tab: initialization strong coupling}
\end{table}
For both optimizations with $\Delta_0 > 0$, the optimization converges to a paired state, with fairly similar correlation functions. In contrast,  for $\Delta_0 = 0$, the optimization finds a state with no pairing that has significantly higher energy. 



\begin{figure*}[t] 
    \centering
    \includegraphics[width=0.9\linewidth]{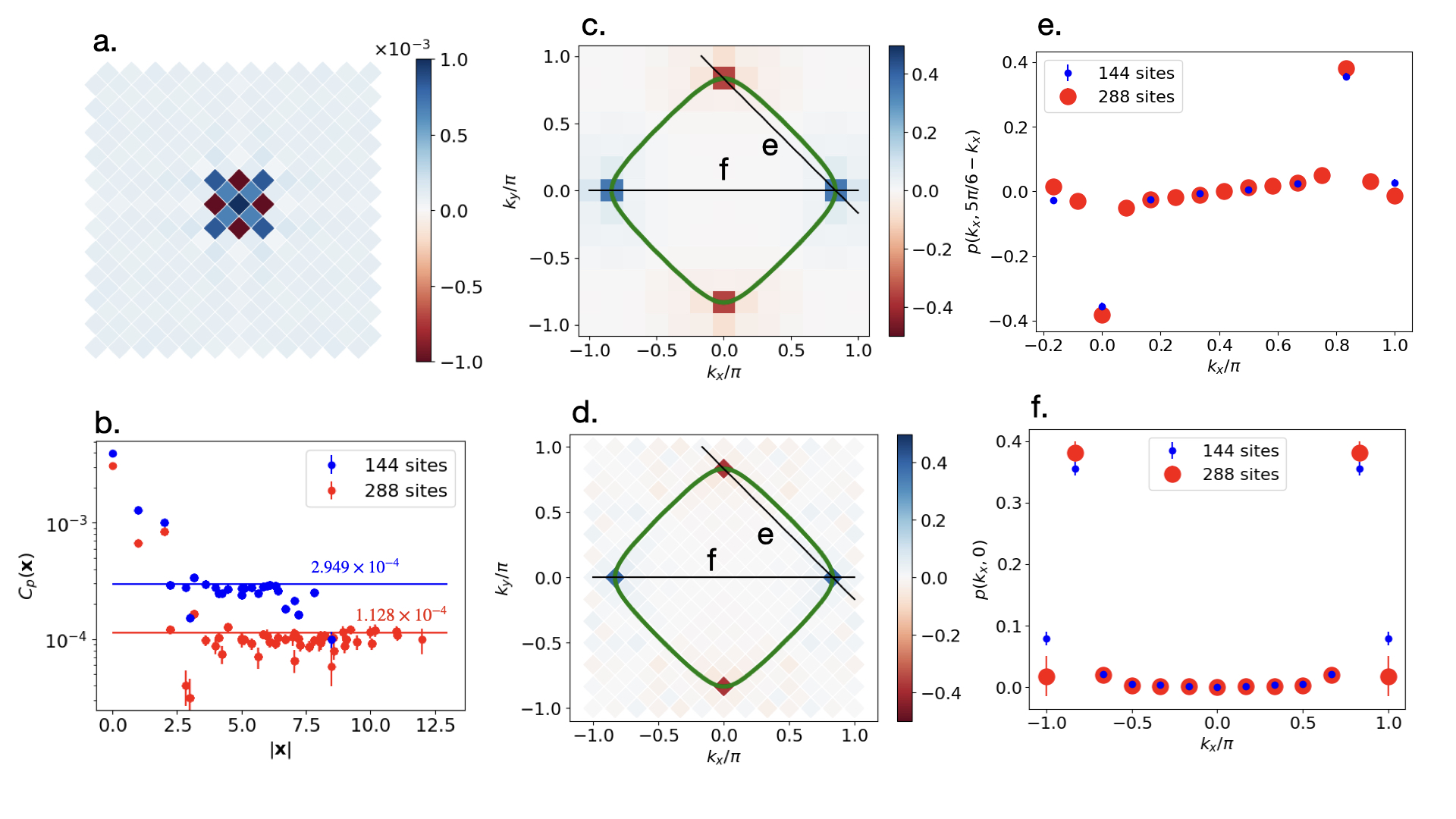}
    \caption{{\bf Long range pairing order in the $t^\prime = 0$ Hubbard model at 1/6 doping}. Results are shown for $144$ and $288$ site simulation clusters for $U=4$, $\delta=1/6$ {\bf a.} Color plot of d-wave pair correlation functions of the $288$ site lattice. {\bf b.} Line plot showing pair correlations at $U=4$ as a function of distance on $144$ (blue) and $288$ (red) site simulation clusters. Lines that denote the average pair correlation, ${\bar C}_p$, on both system sizes are shown for reference. {\bf c.-d.} Color plots of momentum space pairing estimator, $p({\bf k})$ on $144$ ({\bf c.}) site and $288$ site ({\bf d.)} clusters. {\bf e-f.} Momentum space pairing $p({\bf k})$ along the one-dimensional cuts shown in {\bf c.-d.} for the $144$ (blue) and $288$ (red) site clusters. 
    \label{fig: pure hubbard}   
    }
\end{figure*}

\section{Superconductivity in the overdoped Hubbard model} \label{sec: uniform SC}

\subsection{Pair correlation function and finite size effects}

As mentioned in the introduction, a much debated question is whether the pure Hubbard model has a superconducting ground state in any region of the parameter space. This is a relevant question for optical lattices 
which are beginning to reach effective temperatures where superconductivity might be detected\cite{xu_nature_2025}. 

At a formal level, the Hubbard model is known to be a d-wave superconductor for perturbatively small $U$ \cite{KohnLuttinger, Raghu2010weakcoupling}. However, the critical temperature is exponentially small in $\sim (t/U)^2$, 
corresponding to very weak pairing.

An important practical goal is to figure out whether pairing becomes substantial once $U$ is increased beyond the perturbative limit, but not so large as to induce stripe ordering. This regime has been studied using diagrammatic Monte Carlo, which detects a BCS-like pairing instability for $U/t < 4$ \cite{Deng_epl_2015}.

We focus on the point $U=4$, where the stripes are melted \cite{xu2022stripe,peters_2014}, but $U$ is substantially beyond the weak coupling regime. We train HFPS wavefunctions starting from various mean field orbitals $F$, corresponding to a filled stripe, a metal, and d-wave superconductors with various amounts of pairing. All of the initializations use the symmetry breaking cell of the filled stripe phase. The optimization finds two stable local minima; a superconducting phase with d-wave pairing and weak magnetic order, and a filled-stripe phase with no pairing. The superconductor is lower in energy by $\sim 0.5 \%$. The optimization energies on a $144$ site cluster starting from different initial conditions are shown below.  

\begin{table}[H]
    \centering
    \begin{tabular}{|c|c|c|c|c|}
    \hline
          $\Delta_0$ &$U_0$&E/N &${\bar C}_p(|{\bf x}| > 3)$&Phase\\ \hline
           0.0&1&-1.0717(2)&  $2.17(1)  \times 10^{-4}$&SC\\ \hline
 0.2& 3& -1.0737(1)& $3.65(1)  \times 10^{-4}$&SC\\ \hline
 \textbf{0.5}& \textbf{3}& \textbf{-1.0751(1)}&$\bf{2.57(1)  \times 10^{-4}}$&\textbf{SC}\\ \hline
 1.0& 3& -1.0743(1)& $3.19(1) \times 10^{-4}$&SC\\ \hline
 0.0& 3& -1.0700(2)&$6.66(1) \times 10^{-6}$ &Stripe\\ \hline \end{tabular}
    \caption{Energies per site and average pair correlations (for pair separations of a distance larger than 3) for simulations on a $144$ site cluster at $\delta = 1/6$, $U=4$ and $t^\prime=0$ using different initial Pfaffian orbitals. The orbitals are generated from exact gradient descent on the Hamiltonian equation  \ref{eqn: mean field ham} with the parameters specified in the table. The lowest energy solution, which has long range pairing order, is shown in bold.}
    \label{tab:initialization weak coupling}
\end{table}
Even when the mean field orbitals are initialized in a metallic state by choosing $\Delta_0 = 0, U_0 = 1$, the optimization finds a state with a significant amount of pairing. This contrasts with the half-filled stripe discussed in the previous section, where a pairing field was necessary to find the lowest-energy phase.

After verifying that the lowest energy state has superconducting order we examine how pair correlations change with system size. In Fig. \ref{fig: pure hubbard}b., the pair correlation functions (Eqn. \ref{eqn: d-wave pair-correlation}) on $144$ and $288$ site clusters are compared. While the correlations on both clusters asymptote to non-zero values at fairly short distances, the $288$ site cluster decays to a much smaller value. The average pair correlation, ${\bar C}_p$, which should be non-zero for a superconductor in the thermodynamic limit, is displayed for as a horizontal line for reference. This decays by roughly a factor of $3$ when the system size is doubled. 

From this data, it is thus impossible to infer whether ${\bar C}_p$ is finite in the limit of infinite system size. Therefore, a different way to measure pairing is necessary to resolve whether the ground state is a superconductor. We turn to weak coupling d-wave BCS theory \cite{Raghu2010weakcoupling, KohnLuttinger} for inspiration. 
The finite size effects from BCS theory have been studied extensively \cite{anderson1959theory, black1996spectroscopy, von1996parity, golubev1994parity}. In general they predict that pair correlations will decay with system size when the pair correlation length is much larger than the distance around the torus $\hbar v_F/\Delta \gg \sqrt{N}$. In this regime, pairs are fully delocalized on the torus such that they have a roughly equal probability of having each binding radius. As system size is increased, the pairs spread out, and the probability of a pair being local goes like $\mathcal{O}(1/N)$. 
The average pair correlation ${\bar C}_p$,  which measures tunneling processes for \textit{local} pairs, 
displays a $1/N^2$ dependence on system size up to very large system sizes before eventually saturating -- see appendix \ref{sec: finite size BCS} for detailed analysis of d-wave BCS theory on finite size clusters.  

In the following section, we show that at $U=4, \delta = 1/6$ our simulations are still in the regime where the pair correlation length is larger than the distance around the torus. This explains the strong decay in the average pair correlation with system size. By moving to momentum space, we are able to construct a robust measure of pairing with weak finite size effects. 

\subsection{Measuring momentum space pairing} \label{sec: k-space pairing appendix}

Just as we defined a real space pair correlation function in Eqn. \ref{eqn: d-wave pair-correlation}, we can also define a momentum space pair correlation,
\begin{equation*} \label{eqn: momentum pair correlation}
    M_{\bf k \bf k'} = \langle {\hat p}^\dagger_{\bf k} {\hat p}_{\bf k'}   \rangle -  \delta_{\bf k \bf k'} \langle {\hat n}_{\bf k \uparrow} \rangle \langle {\hat n}_{-\bf k \downarrow} \rangle.  
\end{equation*}
Here ${\hat p}_{\bf k} = \langle {\hat c}_{\bf k, \uparrow}  {\hat c}_{\bf -k, \downarrow}\rangle$ annihilates a momentum space pair and ${\hat n}_{\bf k \sigma} = {\hat c}^\dagger_{\bf k , \sigma} {\hat c}_{\bf k, \sigma} $ measures the occupation of a particle with momentum ${\bf k}$. The normal state contribution is subtracted, such that $M_{\bf k \bf k'} = 0$ for free fermions. In contrast, for a uniform superconductor $M_{\bf k \bf k'}$ should be a rank-1 object that describes the independent creation and annihilation of pairs,
\begin{equation} \label{eqn: rank-1 approx}
\begin{cases}
    \lim_{N \rightarrow \infty} M_{\bf k \bf k'} = p({\bf k}) p({\bf k'}) & \textrm{SC} \\  \lim_{N \rightarrow \infty} M_{\bf k \bf k'} = 0  & \textrm{No SC}
    \end{cases}
\end{equation}
Since $M$ involves two momentum space operators, it is a very expensive object for a real space code. Therefore, we only  compute $M$ for a single simulation ($U=4, \delta = 1/6$ on a $144$ cluster). 
In that case, we indeed find that $M$ is rank-1 to good approximation (see. App. \ref{sec: appendix eigenvalue M} for details), which allows us to extract $p({\bf k})$ using a Krylov method. 

If we assume that $M_{\bf k, \bf k'}$ has the rank-1 form in Eqn. \ref{eqn: rank-1 approx}, $p({\bf k})$ can be computed in terms of a mixed real-momentum space correlation function
\begin{equation} \label{eqn: momentum space pairing proxy}
    p({\bf k}) =   \frac{c_{m} ({\bf k})}{(\frac{1}{N}  \sum_{\bf k} g_{\bf k} c_{m} ({\bf k}))^{1/2} } 
\end{equation}
\begin{equation} \label{eqn: mixed estimator}
  c_{m} ({\bf k}) =   \langle {\hat d}^\dagger {\hat p}_{\bf k} \rangle - \frac{1}{N} g_{\bf k} \langle n_{\bf k \uparrow}\rangle \langle n_{\bf -k, \downarrow} \rangle
\end{equation}
where $g_{\bf k} = (\cos(k_x) - \cos(k_y))/2$ is the d-wave form factor. This estimator suppress the noise in $p({\bf k})$ by a factor of $\sqrt{N}$ relative to using the dominant eigenvector of $M_{\bf k \bf k'}$. 

We plot $p({\bf k})$ at $U=4$ on $144$ and $288$ site clusters in Fig. \ref{fig: pure hubbard}c.-d. For both system sizes, $p({\bf k})$ is sharply peaked at the Fermi surface and acquires a sign change under a 
$90 \degree$ rotation.  In Fig. \ref{fig: pure hubbard}e.-f., we plot 1-dimensional cuts of $p({\bf k})$ along relevant directions of the Fermi surface, and find that the value of $p({\bf k})$ is fairly similar across momenta common to both clusters. As seen in figure  \ref{fig: pure hubbard}e. the 288 site cluster has additional k-points which help to resolve the peaks better.  

Crucially, our data show that the momentum space pairing is roughly invariant with system size, in contrast to 
the large-distance behavior of the pair correlations.
This therefore provides strong evidence for superconductivity. 
Since the peak in $p({\bf k})$ is under-resolved by the momentum space grid, the pair correlation length is longer than the maximum spacing on the torus. We can see this explicitly by writing the average pair correlation (equation \ref{eqn: d-wave pair-correlation}) in terms of $p({\bf k})$ as follows
\begin{equation} \label{eqn: p to pair correlation}
    {\bar C}_p = \Bigg|\frac{1}{N} \sum_{\bf k} g_{\bf k} p({\bf k})\Bigg|^2
\end{equation}
Since $p({\bf k})$ is peaked at the anti-node of the Fermi surface, this sum is dominated by a single point leading to a large $1/N^2$ contribution to the scaling with system size, 
hence explaining the distinct behavior of momentum-space and real-space measures of pairing.

\begin{figure*}[t] 
    \centering
    \includegraphics[width=0.9\linewidth]{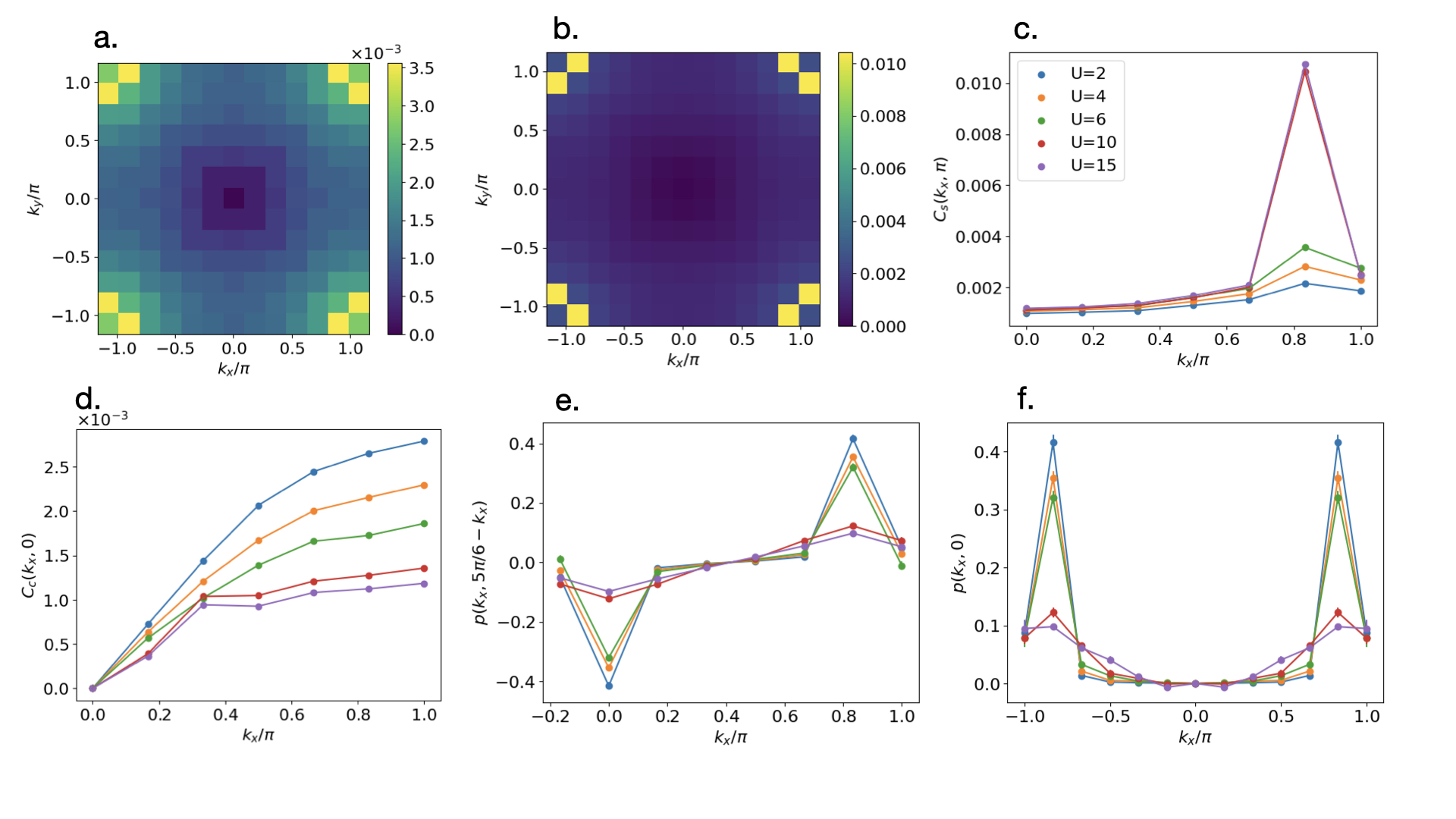}
    \caption{{\bf Correlation functions in the $t^\prime = 0$ Hubbard model for $\boldsymbol{\delta} \bf{= 1/6}$ at different values of $\bf{U}$}. Simulations are shown for different values of $U$ on a $144$ site cluster. {\bf a.} Spin structure factor, $C_s({\bf k})$ for $U=6$ {\bf b.} Spin structure factor at $U=10$ {\bf c.} One-dimensional cut of the spin structure $C_s({\bf k})$ factor along $k_y= \pi$. {\bf d.} One dimensional cut of the charge structure factor $C_c({\bf k})$ along $k_y = 0$. {\bf e.} P({\bf k}) plotted along the same cut in Fig. \ref{fig: pure hubbard}e. {\bf f.} P({\bf k}) plotted along the same cut in Fig. \ref{fig: pure hubbard}f.
    \label{fig: correlation fuctions as U is changed}
    }
\end{figure*}

\subsection{Fate of the uniform superconductor as U is increased} \label{sec: changing U}

So far, our discussion has been limited to the point $U=4$, where the pairs are still highly de-localized. For larger $U$ there is a competing filled-stripe phase, equivalent to the one in \ref{sec: stripy SC}, except with a different period $\lambda_{s(c)} = 12(6)$. A CP-AFQMC calculation, which considered the energetic competition between a 
non-superconducting metal and a stripe, estimates that there is a phase transition around $U \approx 5$ \cite{xu2022stripe}. However, since the superconducting phase gains energy relative to the metal, this transition may be pushed up to larger $U$.  

We use transfer learning to train models at $U=2,4,6,10,15$ on a $144$ site cluster. We consider the lowest energy stripe and superconducting phases specified in table \ref{tab: initialization strong coupling} 
As  $U$ is changed, the superconducting branch is always lowest in energy, but for $U=10,15$ it also develops striped magnetic order. We probe this magnetic order by looking at the charge/spin structure factors,
\begin{equation}
  C_{c(s)}({\bf k}) = \frac{1}{N} \sum_{\bf x} e^{i {\bf k} \cdot {\bf x}} C_{c(s)}({\bf x}).  
\end{equation}
For filled-stripe order, the structure factor peaks are at $k_c = (0,2 \pi/\delta)$ and $k_s = (\pi -\pi/\delta, \pi)$. Fig. \ref{fig: correlation fuctions as U is changed}a.-b., shows the spin structure factor at $U=6$, which has a dispersive peak at $k_s$ and $U=10$, which has a sharp peak at $k_s$. This may indicate a phase transition from a metal to a stripe phase.  In Fig. \ref{fig: correlation fuctions as U is changed}c.-d., we plot one-dimensional cuts of the spin and charge structure factors that bisect these peaks. We see that going from $U=6$ to $U=10$, the spin structure factor peak at $k_s$ becomes much sharper and the charge structure factor develops a kink at $k_c$  . 

We plot relevant one-dimensional cuts of the momentum space pairing, $p({\bf k})$ in Fig \ref{fig: correlation fuctions as U is changed}e.-f. As $U$ is increased the peaks tend to become less sharp, indicating 
a crossover away from the BCS regime. Again, there is a pronounced difference in the correlation function between $U=6$ and $U=10$. Once stripe order sets in, the pairs are de-localized in k-space, corresponding to a shorter correlation length. 

In summary, at hole doping $\delta=1/6$ and $t^\prime=0$, we find evidence for a phase transition from a uniform superconductor to a phase with stripe order between $U=6$ and $U=10$. While this stripe phase still has significant pairing order on a $144$ site cluster, it remains to be seen whether this persists to the thermodynamic limit. 


\section{Discussion}

We have presented a study of superconductivity in the Hubbard model using a variational ansatz that combines Pfaffians with deep neural networks. Using fast Pfaffian algorithms \cite{chen2025pfaffian, Tahara_JPSJ08_mVMC, xu2022optimized}, these simulations are scaled to large symmetric clusters of up to $288$ sites, allowing us to capture the ground state physics with minimal geometric bias. This has given new insight into the behavior of superconductors on finite sized tori. 

We focus on two regimes of the Hubbard model, where the ground state may be a superconductor. First we look at the $t-t^\prime$ model at strong coupling and low-doping with $t^\prime \le 0$, which is an effective model for hole doped cuprate superconductors.  Here, we find that an insulating filled-stripe phase is stable at $t^\prime = 0$, which transitions to a half-filled stripe with superconductivity around $t^\prime = -0.1$.  Using a Pfaffian based ansatz is crucial to represent this phase, as it has both magnetic and superconducting order. 

Our results largely confirm the recent handshake study combining CP-AFQMC and DMRG~\cite{Xu_Science24_Hubbard}, 
which found that partially-filled stripes coexist with superconductivity in the $t^\prime < 0$ Hubbard model. 
Remarkably, this establishes consistency between three distinct computational methods on this challenging question. However, it should be noted that a recent study using DMRG on 6-leg cylinders reaches a different conclusion \cite{jiang_6legs_2024}. 
It is possible that this 
is due to the different geometry and boundary conditions used, which may favor charge- and spin-density ordering (see App. \ref{sec: geometry} for a discussion of geometric effects in our simulations). This obviously deserves further comparative studies.

For the particular $16 \times 16$ cluster we study, the half-filled stripe is favorable, but it is possible that different partially-filled stripe is favored in the thermodynamic limit. From a Hartree-Fock perspective, the filled stripe has a Mott gap while the partially-filled stripes have additional electrons in higher energy bands \cite{sorella2023systematically}. This may explain the difference in pairing between filled and partially-filled stripes. Future work should compare the energies and superconducting correlations among partially-filled stripes of different periods. 

Next we discuss the pure Hubbard model with $t^\prime = 0$, which is known to be a d-wave superconductor in the weak coupling limit for low doping levels\cite{Raghu2010weakcoupling,arovas_ARCMP_2022}. 
We look for superconductivity at intermediate couplings and higher doping\cite{sorella2023systematically}, $U=4$, $\delta = 1/6$, where stripes are unlikely to be energetically competitive. We find that the ground state has short ranged magnetic order and pairing order on finite size clusters. While the correlations for spatially localized d-wave pairs decay like $1/N^2$ with system size, the correlations between momentum space pairs are fairly size invariant. From this data, we conclude that the ground state at $U=4$ is a superconductor with weakly bound pairs, such that the pair correlation length is longer than the circumference of the largest simulation cluster we can access. This data is in strong agreement with the diagrammatic   Monte Carlo calculation in reference \cite{Deng_epl_2015}, which concludes the physics is governed by a BCS-like instability from a re-normalized Fermi-liquid state. We note that the magnitude of the local order parameter that can be inferred from Fig. \ref{fig: pure hubbard}b. is not inconsistent with those reported in reference \cite{Xu_Science24_Hubbard} obtained by linear response to an applied pairing field (see Fig.13 in that reference).      

Our data analysis introduces a novel approach for extracting information about superconductivity from finite size numerical simulations of the repulsive Hubbard model in regimes where the pairs are weakly bound. For the overdoped regime ($\delta = 1/6$), we find that the pairing is better understood in momentum space than real space. Even as $U$ is increased far beyond the BCS regime, the momentum space pairing is still size invariant. This suggests that qualitative features of the weak coupling regime may apply beyond the range where it is controlled.  

Our work establishes that Pfaffian-based NQS is an excellent tool for revealing superconducting order. Future NQS work should aim at further lowering 
the energies of these calculations by improving the optimization algorithms and documenting how the correlation functions change in this process. 

Our ability to simulate large systems opens the possibility of using Pfaffian-based NQS for more realistic models of strongly correlated quantum materials. 
Applications such as three-band models for cuprates \cite{emery1987theory}, as well as multi-orbital and moir\'e materials should be feasible. 


\section*{Acknowledgements}

The NQS simulations are performed using \texttt{Quantax}~\cite{quantax}, in which the scalable ANNs are implemented by \texttt{JAX}~\cite{jax2018github} and \texttt{Equinox}~\cite{Kidger_arxiv21_Equinox}, and the LRU is implemented by \texttt{lrux}~\cite{lrux}. 
We are grateful to Annabelle Bohrdt, Garnet Chan, Fabian Grusdt, Hannah Lange, Ryan Levy, Miguel Morales, Javier Robledo-Moreno, Riccardo Rende, Conor Smith, Steve White and Shiwei Zhang for useful discussions. The Flatiron Institute is a division of the Simons Foundation. We acknowledge the 
help of its Scientific Computing Core in making our computations possible.

\bibliographystyle{apsrev4-2}
\bibliography{references}

@misc{kendrick_2025,
      title={Pseudogap in a Fermi-Hubbard quantum simulator}, 
      author={Lev Haldar Kendrick and Anant Kale and Youqi Gang and Alexander Dennisovich Deters and Martin Lebrat and Aaron W. Young and Markus Greiner},
      year={2025},
      eprint={2509.18075},
      archivePrefix={arXiv},
      primaryClass={cond-mat.quant-gas},
      url={https://arxiv.org/abs/2509.18075}, 
}

@misc{chalopin_2024,
      title={Probing the magnetic origin of the pseudogap using a Fermi-Hubbard quantum simulator}, 
      author={Thomas Chalopin and Petar Bojović and Si Wang and Titus Franz and Aritra Sinha and Zhenjiu Wang and Dominik Bourgund and Johannes Obermeyer and Fabian Grusdt and Annabelle Bohrdt and Lode Pollet and Alexander Wietek and Antoine Georges and Timon Hilker and Immanuel Bloch},
      year={2024},
      eprint={2412.17801},
      archivePrefix={arXiv},
      primaryClass={cond-mat.str-el},
      url={https://arxiv.org/abs/2412.17801}, 
}

@article{Deng_epl_2015,
doi = {10.1209/0295-5075/110/57001},
url = {https://doi.org/10.1209/0295-5075/110/57001},
year = {2015},
month = {jun},
publisher = {EDP Sciences, IOP Publishing and Società Italiana di Fisica},
volume = {110},
number = {5},
pages = {57001},
author = {Deng, Youjin and Kozik, Evgeny and Prokof'ev, Nikolay V. and Svistunov, Boris V.},
title = {Emergent BCS regime of the two-dimensional fermionic Hubbard model: Ground-state phase diagram},
journal = {Europhysics Letters}
}

@article{jiang_6legs_2024,
  title = {Ground-state phase diagram and superconductivity of the doped Hubbard model on six-leg square cylinders},
  author = {Jiang, Yi-Fan and Devereaux, Thomas P. and Jiang, Hong-Chen},
  journal = {Phys. Rev. B},
  volume = {109},
  issue = {8},
  pages = {085121},
  numpages = {6},
  year = {2024},
  month = {Feb},
  publisher = {American Physical Society},
  doi = {10.1103/PhysRevB.109.085121},
  url = {https://link.aps.org/doi/10.1103/PhysRevB.109.085121}
}

@article{zhang_vondelft_2025,
  title = {Frustration-Induced Superconductivity in the $t\text{\ensuremath{-}}{t}^{\ensuremath{'}}$ Hubbard Model},
  author = {Zhang, Changkai and Li, Jheng-Wei and Nikolaidou, Dimitra and von Delft, Jan},
  journal = {Phys. Rev. Lett.},
  volume = {134},
  issue = {11},
  pages = {116502},
  numpages = {7},
  year = {2025},
  month = {Mar},
  publisher = {American Physical Society},
  doi = {10.1103/PhysRevLett.134.116502},
  url = {https://link.aps.org/doi/10.1103/PhysRevLett.134.116502}
}

@article{kancharla_2008,
  title = {Anomalous superconductivity and its competition with antiferromagnetism in doped Mott insulators},
  author = {Kancharla, S. S. and Kyung, B. and S\'en\'echal, D. and Civelli, M. and Capone, M. and Kotliar, G. and Tremblay, A.-M. S.},
  journal = {Phys. Rev. B},
  volume = {77},
  issue = {18},
  pages = {184516},
  numpages = {12},
  year = {2008},
  month = {May},
  publisher = {American Physical Society},
  doi = {10.1103/PhysRevB.77.184516},
  url = {https://link.aps.org/doi/10.1103/PhysRevB.77.184516}
}

@article{gullmillis_natphys_2015,
  title = {Numerical models come of age},
  volume = {11},
  ISSN = {1745-2481},
  url = {http://dx.doi.org/10.1038/nphys3501},
  DOI = {10.1038/nphys3501},
  number = {10},
  journal = {Nature Physics},
  publisher = {Springer Science and Business Media LLC},
  author = {Gull,  E. and Millis,  A. J.},
  year = {2015},
  month = oct,
  pages = {808–810}
}

@article{xu_nature_2025,
  title = {A neutral-atom Hubbard quantum simulator in the cryogenic regime},
  volume = {642},
  ISSN = {1476-4687},
  url = {http://dx.doi.org/10.1038/s41586-025-09112-w},
  DOI = {10.1038/s41586-025-09112-w},
  number = {8069},
  journal = {Nature},
  publisher = {Springer Science and Business Media LLC},
  author = {Xu,  Muqing and Kendrick,  Lev Haldar and Kale,  Anant and Gang,  Youqi and Feng,  Chunhan and Zhang,  Shiwei and Young,  Aaron W. and Lebrat,  Martin and Greiner,  Markus},
  year = {2025},
  month = jun,
  pages = {909–915}
}

@article{arovas_ARCMP_2022,
   author = "Arovas, Daniel P. and Berg, Erez and Kivelson, Steven A. and Raghu, Srinivas",
   title = "The Hubbard Model", 
   journal= "Annual Review of Condensed Matter Physics",
   year = "2022",
   volume = "13",
   number = "Volume 13, 2022",
   pages = "239-274",
   doi = "https://doi.org/10.1146/annurev-conmatphys-031620-102024",
   url = "https://www.annualreviews.org/content/journals/10.1146/annurev-conmatphys-031620-102024",
   publisher = "Annual Reviews",
   issn = "1947-5462",
   type = "Journal Article"
  }

@article{DopingMottInsulator,
  title = {Doping a Mott insulator: Physics of high-temperature superconductivity},
  author = {Lee, Patrick A. and Nagaosa, Naoto and Wen, Xiao-Gang},
  journal = {Rev. Mod. Phys.},
  volume = {78},
  issue = {1},
  pages = {17--85},
  numpages = {0},
  year = {2006},
  month = {Jan},
  publisher = {American Physical Society},
  doi = {10.1103/RevModPhys.78.17},
  url = {https://link.aps.org/doi/10.1103/RevModPhys.78.17}
}

@article{KotliarSuperexchange,
  title = {Superexchange mechanism and d-wave superconductivity},
  author = {Kotliar, Gabriel and Liu, Jialin},
  journal = {Phys. Rev. B},
  volume = {38},
  issue = {7},
  pages = {5142--5145},
  numpages = {0},
  year = {1988},
  month = {Sep},
  publisher = {American Physical Society},
  doi = {10.1103/PhysRevB.38.5142},
  url = {https://link.aps.org/doi/10.1103/PhysRevB.38.5142}
}

@article{sorella2023systematically,
  title={Systematically improvable mean-field variational ansatz for strongly correlated systems: Application to the Hubbard model},
  author={Sorella, Sandro},
  journal={Physical Review B},
  volume={107},
  number={11},
  pages={115133},
  year={2023},
  publisher={APS}
}

@article{zheng2017stripe,
  title={Stripe order in the underdoped region of the two-dimensional Hubbard model},
  author={Zheng, Bo-Xiao and Chung, Chia-Min and Corboz, Philippe and Ehlers, Georg and Qin, Ming-Pu and Noack, Reinhard M and Shi, Hao and White, Steven R and Zhang, Shiwei and Chan, Garnet Kin-Lic},
  journal={Science},
  volume={358},
  number={6367},
  pages={1155--1160},
  year={2017},
  publisher={American Association for the Advancement of Science},
url = {https://www.science.org/doi/10.1126/science.aam7127}
}

@article{Anderson_Science87_RVB,
  title={The resonating valence bond state in {L}a$_2${C}u{O}$_4$ and superconductivity},
  author={Anderson, Philip W},
  journal={Science},
  volume={235},
  number={4793},
  pages={1196--1198},
  year={1987},
  publisher={American Association for the Advancement of Science},
url = {https://www.science.org/doi/10.1126/science.235.4793.1196}
}

@article{imadaSCpurehubbard,
  title = {Stripe and superconducting order competing in the Hubbard model on a square lattice studied by a combined variational Monte Carlo and tensor network method},
  author = {Darmawan, Andrew S. and Nomura, Yusuke and Yamaji, Youhei and Imada, Masatoshi},
  journal = {Phys. Rev. B},
  volume = {98},
  issue = {20},
  pages = {205132},
  numpages = {11},
  year = {2018},
  month = {Nov},
  publisher = {American Physical Society},
  doi = {10.1103/PhysRevB.98.205132},
  url = {https://link.aps.org/doi/10.1103/PhysRevB.98.205132}
}

@article{huang2018stripe,
  title={Stripe order from the perspective of the Hubbard model},
  author={Huang, Edwin W and Mendl, Christian B and Jiang, Hong-Chen and Moritz, Brian and Devereaux, Thomas P},
  journal={npj Quantum Materials},
  volume={3},
  number={1},
  pages={22},
  year={2018},
  publisher={Nature Publishing Group UK London}
}

@article{emery1987theory,
  title={Theory of high-T c superconductivity in oxides},
  author={Emery, VJ},
  journal={Physical Review Letters},
  volume={58},
  number={26},
  pages={2794},
  year={1987},
  publisher={APS}
}

@article{smith2024unified,
  title={Unified variational approach description of ground-state phases of the two-dimensional electron gas},
  author={Smith, Conor and Chen, Yixiao and Levy, Ryan and Yang, Yubo and Morales, Miguel A and Zhang, Shiwei},
  journal={Physical Review Letters},
  volume={133},
  number={26},
  pages={266504},
  year={2024},
  publisher={APS}
}

@article{xu2022optimized,
  title={Optimized implementation for calculation and fast-update of Pfaffians installed to the open-source fermionic variational solver mVMC},
  author={Xu, RuQing G and Okubo, Tsuyoshi and Todo, Synge and Imada, Masatoshi},
  journal={Computer Physics Communications},
  volume={277},
  pages={108375},
  year={2022},
  publisher={Elsevier}
}

@article{Tahara_JPSJ08_mVMC,
author = {Tahara ,Daisuke and Imada ,Masatoshi},
title = {Variational Monte Carlo Method Combined with Quantum-Number Projection and Multi-Variable Optimization},
journal = {Journal of the Physical Society of Japan},
volume = {77},
number = {11},
pages = {114701},
year = {2008},
doi = {10.1143/JPSJ.77.114701},
URL = {https://doi.org/10.1143/JPSJ.77.114701},
}

@article{Ruckenstein_mean_field_J,
  title = {Mean-field theory of high-${T}_{c}$ superconductivity: The superexchange mechanism},
  author = {Ruckenstein, Andrei E. and Hirschfeld, Peter J. and Appel, J.},
  journal = {Phys. Rev. B},
  volume = {36},
  issue = {1},
  pages = {857--860},
  numpages = {0},
  year = {1987},
  month = {Jul},
  publisher = {American Physical Society},
  doi = {10.1103/PhysRevB.36.857},
  url = {https://link.aps.org/doi/10.1103/PhysRevB.36.857}
}

@article{VMC_Hubbard_tprime_Lechiara,
  title={Variational Monte Carlo study of stripes as a function of doping in the $ tt\^{}$\{$\^{}$\{$$\backslash$prime$\}$$\}$ $ Hubbard model},
  author={Lechiara, Antonio and Marino, Vito and Tocchio, Luca F},
  journal={Journal of Physics: Condensed Matter},
  volume={36},
  number={39},
  pages={395602},
  year={2024},
  publisher={IOP Publishing}
}

@article{DQMC_Chen,
  title = {Enhancement of dominant ${d}_{xy}$-wave pairing in the overdoped Hubbard model with next-nearest hopping ${t}^{\ensuremath{'}}$},
  author = {Chen, Chao and Chen, Cong and Li, Kai and Sui, Xuelei and Long, Wenbo and Yang, Haibin and Huang, Bing},
  journal = {Phys. Rev. B},
  volume = {112},
  issue = {11},
  pages = {115148},
  numpages = {7},
  year = {2025},
  month = {Sep},
  publisher = {American Physical Society},
  doi = {10.1103/wfbr-fg2v},
  url = {https://link.aps.org/doi/10.1103/wfbr-fg2v}
}

@InProceedings{Cohen_arxiv16_GCNN,
  title = 	 {Group Equivariant Convolutional Networks},
  author = 	 {Cohen, Taco and Welling, Max},
  booktitle = 	 {Proceedings of The 33rd International Conference on Machine Learning},
  pages = 	 {2990--2999},
  year = 	 {2016},
  editor = 	 {Balcan, Maria Florina and Weinberger, Kilian Q.},
  volume = 	 {48},
  series = 	 {Proceedings of Machine Learning Research},
  address = 	 {New York, New York, USA},
  month = 	 {20--22 Jun},
  publisher =    {PMLR},
  url = 	 {https://proceedings.mlr.press/v48/cohenc16.html},
}

@article{Astrakhantsev_PRX21_pyrochlore,
  title = {Broken-Symmetry Ground States of the Heisenberg Model on the Pyrochlore Lattice},
  author = {Astrakhantsev, Nikita and Westerhout, Tom and Tiwari, Apoorv and Choo, Kenny and Chen, Ao and Fischer, Mark H. and Carleo, Giuseppe and Neupert, Titus},
  journal = {Phys. Rev. X},
  volume = {11},
  issue = {4},
  pages = {041021},
  numpages = {20},
  year = {2021},
  month = {Oct},
  publisher = {American Physical Society},
  doi = {10.1103/PhysRevX.11.041021},
  url = {https://link.aps.org/doi/10.1103/PhysRevX.11.041021}
}

@article{Moreno_PNAS22_HFDS,
    author = {Javier Robledo Moreno  and Giuseppe Carleo  and Antoine Georges  and James Stokes },
    title = {Fermionic wave functions from neural-network constrained hidden states},
    journal = {Proceedings of the National Academy of Sciences},
    volume = {119},
    number = {32},
    pages = {e2122059119},
    year = {2022},
    doi = {10.1073/pnas.2122059119},
    URL = {https://www.pnas.org/doi/abs/10.1073/pnas.2122059119},
}

@article{Roth_PRB23_GCNN,
  title = {High-accuracy variational Monte Carlo for frustrated magnets with deep neural networks},
  author = {Roth, Christopher and Szab\'o, Attila and MacDonald, Allan H.},
  journal = {Phys. Rev. B},
  volume = {108},
  issue = {5},
  pages = {054410},
  numpages = {12},
  year = {2023},
  month = {Aug},
  publisher = {American Physical Society},
  doi = {10.1103/PhysRevB.108.054410},
  url = {https://link.aps.org/doi/10.1103/PhysRevB.108.054410}
}

@Article{Rende_CP24_SRt,
    author={Rende, Riccardo
    and Viteritti, Luciano Loris
    and Bardone, Lorenzo
    and Becca, Federico
    and Goldt, Sebastian},
    title={A simple linear algebra identity to optimize large-scale neural network quantum states},
    journal={Communications Physics},
    year={2024},
    month={Aug},
    day={02},
    volume={7},
    number={1},
    pages={260},
    issn={2399-3650},
    doi={10.1038/s42005-024-01732-4},
    url={https://doi.org/10.1038/s42005-024-01732-4}
}

@article{Chen_NP24_MinSR,
    author={Chen, Ao and Heyl, Markus},
    title={Empowering deep neural quantum states through efficient optimization},
    journal={Nature Physics},
    year={2024},
    month={Sep},
    day={01},
    volume={20},
    number={9},
    pages={1476-1481},
    issn={1745-2481},
    doi={10.1038/s41567-024-02566-1},
    url={https://doi.org/10.1038/s41567-024-02566-1}
}

@article{Xu_Science24_Hubbard,
    author = {Hao Xu  and Chia-Min Chung  and Mingpu Qin  and Ulrich Schollwöck  and Steven R. White  and Shiwei Zhang },
    title = {Coexistence of superconductivity with partially filled stripes in the Hubbard model},
    journal = {Science},
    volume = {384},
    number = {6696},
    pages = {eadh7691},
    year = {2024},
    doi = {10.1126/science.adh7691},
    URL = {https://www.science.org/doi/abs/10.1126/science.adh7691},
}

@article{Jiang_Science19_DMRG-Hubbard,
author = {Hong-Chen Jiang  and Thomas P. Devereaux },
title = {Superconductivity in the doped Hubbard model and its interplay with next-nearest hopping},
journal = {Science},
volume = {365},
number = {6460},
pages = {1424-1428},
year = {2019},
doi = {10.1126/science.aal5304},
URL = {https://www.science.org/doi/abs/10.1126/science.aal5304},
}

@article{Qin_PRX20_Hubbard,
  title = {Absence of Superconductivity in the Pure Two-Dimensional Hubbard Model},
  author = {Qin, Mingpu and Chung, Chia-Min and Shi, Hao and Vitali, Ettore and Hubig, Claudius and Schollw\"ock, Ulrich and White, Steven R. and Zhang, Shiwei},
  collaboration = {Simons Collaboration on the Many-Electron Problem},
  journal = {Phys. Rev. X},
  volume = {10},
  issue = {3},
  pages = {031016},
  numpages = {18},
  year = {2020},
  month = {Jul},
  publisher = {American Physical Society},
  doi = {10.1103/PhysRevX.10.031016},
  url = {https://link.aps.org/doi/10.1103/PhysRevX.10.031016}
}

@article{Qin_ARCMP22_Hubbard,
   author = "Qin, Mingpu and Schäfer, Thomas and Andergassen, Sabine and Corboz, Philippe and Gull, Emanuel",
   title = "The Hubbard Model: A Computational Perspective", 
   journal= "Annual Review of Condensed Matter Physics",
   year = "2022",
   volume = "13",
   number = "Volume 13, 2022",
   pages = "275-302",
   doi = "https://doi.org/10.1146/annurev-conmatphys-090921-033948",
   url = "https://www.annualreviews.org/content/journals/10.1146/annurev-conmatphys-090921-033948",
   publisher = "Annual Reviews",
   issn = "1947-5462",
   type = "Journal Article",
}

@article{KohnLuttinger,
  title = {New Mechanism for Superconductivity},
  author = {Kohn, W. and Luttinger, J. M.},
  journal = {Phys. Rev. Lett.},
  volume = {15},
  issue = {12},
  pages = {524--526},
  numpages = {0},
  year = {1965},
  month = {Sep},
  publisher = {American Physical Society},
  doi = {10.1103/PhysRevLett.15.524},
  url = {https://link.aps.org/doi/10.1103/PhysRevLett.15.524}
}

@article{
simons_colab_stripes_zheng,
author = {Bo-Xiao Zheng  and Chia-Min Chung  and Philippe Corboz  and Georg Ehlers  and Ming-Pu Qin  and Reinhard M. Noack  and Hao Shi  and Steven R. White  and Shiwei Zhang  and Garnet Kin-Lic Chan },
title = {Stripe order in the underdoped region of the two-dimensional Hubbard model},
journal = {Science},
volume = {358},
number = {6367},
pages = {1155-1160},
year = {2017},
doi = {10.1126/science.aam7127},
URL = {https://www.science.org/doi/abs/10.1126/science.aam7127},
eprint = {https://www.science.org/doi/pdf/10.1126/science.aam7127}
}

@misc{chen2025pfaffian,
      title={Neural Network-Augmented Pfaffian Wave-functions for Scalable Simulations of Interacting Fermions}, 
      author={Ao Chen and Zhou-Quan Wan and Anirvan Sengupta and Antoine Georges and Christopher Roth},
      year={2025},
      eprint={2507.10705},
      archivePrefix={arXiv},
      primaryClass={cond-mat.str-el},
      url={https://arxiv.org/abs/2507.10705} 
}

@article{pavarini2001band,
  title={Band-structure trend in hole-doped cuprates and correlation with T c max},
  author={Pavarini, Eva and Dasgupta, I and Saha-Dasgupta, T and Jepsen, O and Andersen, OK},
  journal={Physical review letters},
  volume={87},
  number={4},
  pages={047003},
  year={2001},
  publisher={APS}
}

@misc{Gu_arxiv25_NQS-Hubbard,
      title={Solving the Hubbard model with Neural Quantum States}, 
      author={Yuntian Gu and Wenrui Li and Heng Lin and Bo Zhan and Ruichen Li and Yifei Huang and Di He and Yantao Wu and Tao Xiang and Mingpu Qin and Liwei Wang and Dingshun Lv},
      year={2025},
      eprint={2507.02644},
      archivePrefix={arXiv},
      primaryClass={cond-mat.str-el},
      url={https://arxiv.org/abs/2507.02644}, 
}

@misc{Kidger_arxiv21_Equinox,
      title={Equinox: neural networks in JAX via callable PyTrees and filtered transformations}, 
      author={Patrick Kidger and Cristian Garcia},
      year={2021},
      eprint={2111.00254},
      archivePrefix={arXiv},
      primaryClass={cs.LG},
      url={https://arxiv.org/abs/2111.00254}, 
}

@article{Jiang4leg,
  title = {Ground state phase diagram of the doped Hubbard model on the four-leg cylinder},
  author = {Jiang, Yi-Fan and Zaanen, Jan and Devereaux, Thomas P. and Jiang, Hong-Chen},
  journal = {Phys. Rev. Res.},
  volume = {2},
  issue = {3},
  pages = {033073},
  numpages = {14},
  year = {2020},
  month = {Jul},
  publisher = {American Physical Society},
  doi = {10.1103/PhysRevResearch.2.033073},
  url = {https://link.aps.org/doi/10.1103/PhysRevResearch.2.033073}
}

@misc{jax2018github,
  author = {James Bradbury and Roy Frostig and Peter Hawkins and Matthew James Johnson and Chris Leary and Dougal Maclaurin and George Necula and Adam Paszke and Jake Vander{P}las and Skye Wanderman-{M}ilne and Qiao Zhang},
  title = {{JAX}: composable transformations of {P}ython+{N}um{P}y programs},
  url = {http://github.com/jax-ml/jax},
  version = {0.3.13},
  year = {2018},
}

@misc{quantax,
  author = {Ao Chen and Christopher Roth},
  title = {Quantax: Flexible neural quantum states based on {QuSpin}, {JAX}, and {Equinox}},
  url = {https://github.com/ChenAo-Phys/quantax},
  year = {2025},
}

@misc{lrux,
  author = {Ao Chen and Christopher Roth},
  title = {Fast low-rank update of matrix determinants and pfaffians in {JAX}},
  url = {https://github.com/ChenAo-Phys/lrux},
  year = {2025},
}

@article{Raghu2010weakcoupling,
  title = {Superconductivity in the repulsive Hubbard model: An asymptotically exact weak-coupling solution},
  author = {Raghu, S. and Kivelson, S. A. and Scalapino, D. J.},
  journal = {Phys. Rev. B},
  volume = {81},
  issue = {22},
  pages = {224505},
  numpages = {17},
  year = {2010},
  month = {Jun},
  publisher = {American Physical Society},
  doi = {10.1103/PhysRevB.81.224505},
  url = {https://link.aps.org/doi/10.1103/PhysRevB.81.224505}
}

@article{xu2022stripe,
  title = {Stripes and spin-density waves in the doped two-dimensional Hubbard model: Ground state phase diagram},
  author = {Xu, Hao and Shi, Hao and Vitali, Ettore and Qin, Mingpu and Zhang, Shiwei},
  journal = {Phys. Rev. Res.},
  volume = {4},
  issue = {1},
  pages = {013239},
  numpages = {10},
  year = {2022},
  month = {Mar},
  publisher = {American Physical Society},
  doi = {10.1103/PhysRevResearch.4.013239},
  url = {https://link.aps.org/doi/10.1103/PhysRevResearch.4.013239}
}

@article{peters_2014,
  title = {Spin density waves in the Hubbard model: A DMFT approach},
  author = {Peters, Robert and Kawakami, Norio},
  journal = {Phys. Rev. B},
  volume = {89},
  issue = {15},
  pages = {155134},
  numpages = {8},
  year = {2014},
  month = {Apr},
  publisher = {American Physical Society},
  doi = {10.1103/PhysRevB.89.155134},
  url = {https://link.aps.org/doi/10.1103/PhysRevB.89.155134}
}

@article{anderson1959theory,
  title={Theory of dirty superconductors},
  author={Anderson, Philip W},
  journal={Journal of Physics and Chemistry of Solids},
  volume={11},
  number={1-2},
  pages={26--30},
  year={1959},
  publisher={Elsevier}
}

@article{black1996spectroscopy,
  title={Spectroscopy of the superconducting gap in individual nanometer-scale aluminum particles},
  author={Black, CT and Ralph, DC and Tinkham, M},
  journal={Physical review letters},
  volume={76},
  number={4},
  pages={688},
  year={1996},
  publisher={APS}
}

@article{von1996parity,
  title={Parity-affected superconductivity in ultrasmall metallic grains},
  author={von Delft, Jan and Zaikin, Andrei D and Golubev, Dmitrii S and Tichy, Wolfgang},
  journal={Physical review letters},
  volume={77},
  number={15},
  pages={3189},
  year={1996},
  publisher={APS}
}

@article{golubev1994parity,
  title={Parity effect and thermodynamics of canonical superconducting ensembles},
  author={Golubev, DS and Zaikin, AD},
  journal={Physics Letters A},
  volume={195},
  number={5-6},
  pages={380--388},
  year={1994},
  publisher={Elsevier}
}

\clearpage

\appendix


\section{Efficiently measuring momentum space pairing} \label{sec: appendix eigenvalue M}

\begin{figure*}[t] 
    \centering
    \includegraphics[width=0.9\linewidth]{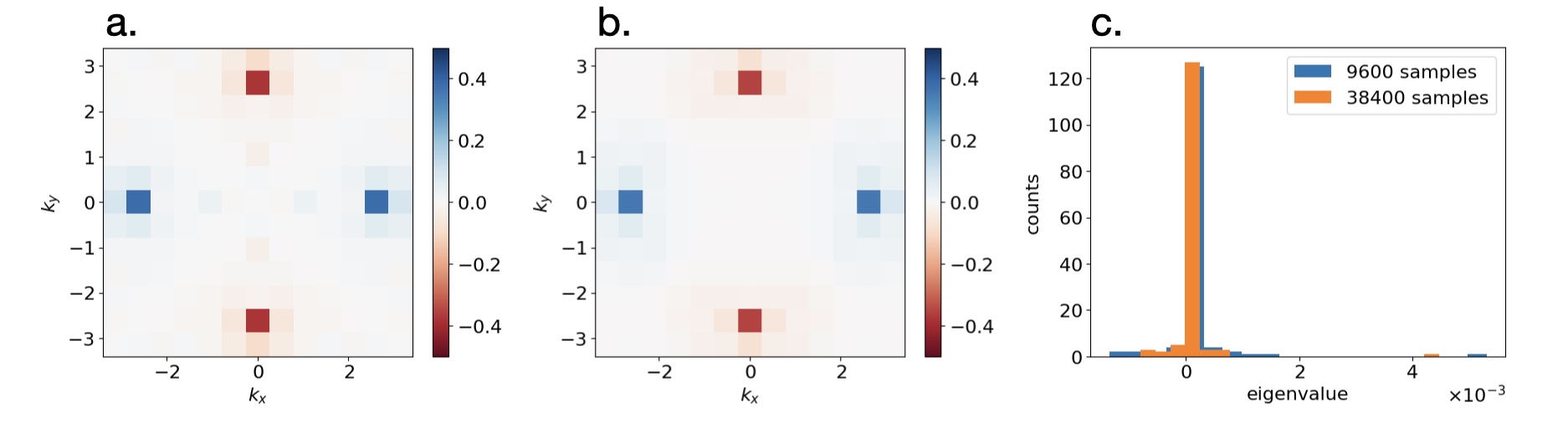}
    \caption{Comparison between methods of computing momentum space pairing at $U=4, \delta = 1/6$ on the 144 site cluster. {\bf a.} Momentum space pairing estimator using dominant eigenmode of $M_{\bf k \bf k'}$, which is computed as $\sqrt{\lambda_{0}} e_0({\bf k})$ (see Eqn. \ref{eqn: momentum space low rank}). {\bf b.} Momentum space pairing,  $p({\bf k})$ using the mixed estimator (Eqn \ref{eqn: momentum space pairing proxy}) {\bf c.} Histogram comparing the eigenvalues of $M_{\bf k, \bf k'}$ using two different sample sizes.}
    \label{fig: kpairingextraction}
\end{figure*}

In the main text (equation \ref{eqn: momentum pair correlation}) we define a momentum space pair correlation function $M_{\bf k \bf k'}$ which examines the correlations between creating and annihilating momentum space cooper pairs. While this correlation is prohibitively expensive for our largest system sizes, we examine its eigenvalue decomposition on a $144$ site cluster at $U=4$,  
\begin{equation} \label{eqn: momentum space low rank}
    M_{\bf k \bf k'} = \sum_i \lambda_i e_i({\bf k}) e_i({\bf k'}).
\end{equation}
We plot the eigenvalues, $\lambda_i$ in Fig. \ref{fig: kpairingextraction}c. and see that there is a single dominant mode, as well as a distribution of subdominant eigenvalues. As we increase the number of samples used to compute $M_{\bf k \bf k'}$, the distribution of the subdominant eigenvalues narrows, indicating they are primarily due to noise. From this we include that, $M_{\bf k \bf k'}$ is already close to rank-1 on our 144 site system. Assuming that $M_{\bf k \bf k'}$ is a rank-1 object with statistical noise,
\begin{equation}
    M_{\bf k \bf k'} = p({\bf k}) p({\bf k'}) + \mathcal{N}(0,\sigma^2),
\end{equation}
we consider the action of $M_{\bf k \bf k'}$ on the d-wave symmetric form factor,
\begin{equation}
  \frac{1}{N} \sum_{\bf k'} M_{\bf k \bf k'} g_{\bf k'} = p({\bf k}) \Big(\frac{{\bf p} \cdot {\bf g}}{N} \Big) + \mathcal{N}(0,\sigma^2/N).
\end{equation}

Here, ${\bf p} \cdot \bf{g}$ measures the overlap between $p({\bf k})$ and the form factor $g_{\bf k}$. Acting $M$ on the form factor extracts the shape of $p({\bf k})$ and reduces the noise by $\sqrt{N}$. In order to extract the norm, we multiply $M$ by the form factor on both sides,

\begin{equation}
  \frac{1}{N^2} \sum_{\bf k, \bf k'} g_{\bf k} M_{\bf k \bf k'} g_{\bf k'} = \Big(\frac{{\bf p} \cdot {\bf g}}{N} \Big)^2 + \mathcal{N}(0,\sigma^2/N^2)
\end{equation}

Using equation \ref{eqn: mixed estimator} we can express $M {\bf g}$ and ${\bf g} M {\bf g}$ in terms of a mixed estimator,
\begin{equation}
 \frac{1}{N} \sum_{\bf k'} M_{\bf k \bf k'} g_{\bf k'} =   \langle {\hat d}^\dagger {\hat p}_{\bf k} \rangle - \frac{1}{N} g_{\bf k} \langle n_{\bf k \uparrow}\rangle \langle n_{\bf -k, \downarrow} \rangle = c_{m} ({\bf k})  
\end{equation}
\begin{equation}
 \frac{1}{N^2} \sum_{\bf k'} g_{\bf k} M_{\bf k \bf k'} g_{\bf k'} = \frac{1}{N} \sum_{\bf k} g_{\bf k} c_{m} ({\bf k})  
\end{equation}

This brings us to equation \ref{eqn: momentum space pairing proxy}, which estimates $p({\bf k})$ in terms of a mixed real/momentum space correlation function. We can see that computing $p({\bf k})$ from the mixed estimator has $\sim \sigma/\sqrt{N}$ noise whereas the largest eigenvalue of $M_{\bf k \bf k'}$ has $\sim \sigma$ noise. We compare the two measures of pairing in figure \ref{fig: kpairingextraction}a.-b., and show they give similar results.

\section{Understanding pair correlations at small U from d-wave BCS theory} \label{sec: finite size BCS}  

In this section we provide a brief introduction to d-wave BCS theory, which can describe the superconductor phenomenologically for small $U$. The dominant finite size effects can be understood by considering the mean field theory with a finite number of k-points.   

\subsection{A brief pedagogical description of BCS theory}

The d-wave BCS Hamiltonian, which perturbs free fermions with a small attractive interaction of d-wave symmetry, is written as follows, 

\begin{equation} \label{eqn: BCS Hamiltonian 1}
    \mathcal{H} = \sum_{\sigma} \sum_{\bf k}  (\epsilon_k - \mu) {\hat n}_{\bf k} + \sum_{\bf k \bf k'} V_{{\bf k}, {\bf k'}} {\hat p}^\dagger_{\bf k} {\hat p}_{\bf k'}  
\end{equation}
where,
\begin{equation} \label{eqn: seperable interaction}
V_{{\bf k}, {\bf k'}} = -\frac{V}{N} g_{\bf k} g_{\bf k'}
\end{equation}
We apply a mean field decoupling to solve this Hamiltonian
\begin{equation} \label{eqn: BCS mean field}
    \mathcal{H}_{BCS} = \sum_{\bf k} \Bigg[ \sum_{\sigma}   (\epsilon_k - \mu) {\hat n}_{\bf k, \sigma} 
    -\big(\Delta_{\bf k} {\hat p}_{\bf k}^\dagger + \textrm{H.C.} \big) \Bigg]
\end{equation}
where 
\begin{equation}
    \Delta_{\bf k} = -\sum_{\bf k'}  V_{{\bf k k'}} \langle {\hat p}_{\bf k'} \rangle.  
\end{equation}
While the true BCS Hamiltonian in equation \ref{eqn: BCS Hamiltonian 1} conserves particle number, the mean field treatment breaks this symmetry and allows for an anomalous expectation value of $\langle p_{\bf k} \rangle$.

Using the separable form in equation \ref{eqn: seperable interaction}, we see that the self-consistent field is  proportional to the form factor,  $\Delta_{\bf k} = \Delta g_{\bf k}$ where 
\begin{equation}
    \Delta = \frac{V}{N} \sum_{\bf k'}  g_{\bf k'} \langle {\hat p}_{\bf k'}  \rangle
\end{equation}
Solving the bilinear Hamiltonian in equation \ref{eqn: BCS mean field} gives the solution
\begin{equation} \label{eqn: momentum space pairing}
\langle {\hat p}_{\bf k} \rangle = \frac{\Delta_{\bf k}}{2((\epsilon_{\bf k} - \mu)^2 + \Delta_{\bf k}^2)^{1/2}}
\end{equation}
Combining these equations leads to the gap equation which relates $\Delta$ to V,
\begin{equation} \label{eqn: gap equation}
    1 = \frac{V}{N} \sum_{\bf k} \frac{g_{\bf k}^2}{2((\epsilon_{\bf k} - \mu)^2 + \Delta^2 g_{\bf k}^2)^{1/2}}  
\end{equation}
The chemical potential is set by the number equation, 

\begin{equation} \label{eqn: number equation}
    n = \frac{1}{2N} \sum_{\bf k} \Bigg( 1 - \frac{\epsilon_{\bf k}-\mu}{((\epsilon_{\bf k} - \mu)^2 + \Delta^2 g_{\bf k})^2)^{1/2}}  \Bigg)
\end{equation}
We can also compute the traditional d-wave order parameter from BCS theory 
\begin{equation} \label{eqn: order param from momentum pairing}
    \langle d \rangle = \frac{1}{N} \sum_{\bf k} g_{\bf k} \langle {\hat p}_{\bf k} \rangle
\end{equation}
\subsection{Fitting pair correlations to BCS theory}

In this section, we fit our momentum space pairing estimator $p({\bf k})$ at $U=2$ to the predictions of mean field BCS theory. In BCS theory, the momentum space pair creation and annihilation processes are fully decoupled. For example, 
\begin{equation}
  \langle {\hat p}^\dagger_{\bf k} {\hat p}_{\bf k'} \rangle =  \langle {\hat p}_{\bf k} \rangle \langle {\hat p}_{\bf k'} \rangle + \delta_{\bf k, \bf k'} \langle {\hat n}_{\bf k, \uparrow} \rangle \langle {\hat n}_{\bf -k, \downarrow} \rangle 
\end{equation}
which gives
\begin{equation}
    M_{\bf k \bf k'} = \langle {\hat p}_{\bf k} \rangle \langle {\hat p}_{\bf k'} \rangle
\end{equation}
Therefore, we expect $M_{\bf k, \bf k'}$ to become low rank as we approach the BCS theory limit. Furthermore, we can compute the mixed estimator
\begin{equation}
    p({\bf k}) = \langle \hat p_{\bf k} \rangle
\end{equation}
and finally the pair-pair correlation function
\begin{equation} \label{eqn: pair pair BCS}
  {\bar C}_p = \frac{1}{N^2}\Big( \sum_{\bf k \bf k'} \langle {\hat p}_{\bf k} \rangle g_{\bf k} \Big)^2
\end{equation}
The finite size effects of ${\bar C}_p$ can be divided into two categories, 
\begin{itemize}
    \item The amplitude and shape of $p(\bf k)$. 
    The peak location is specified by the chemical potential $\mu$, which is set by the equation \ref{eqn: number equation}. The scale is set by $\Delta$
    \item The sum in equation  \ref{eqn: pair pair BCS} strongly depends on the location of allowed k-points in the Brillouin zone. 
\end{itemize}

In the main text, we compare two system sizes, where the chemical potentials are exactly the same for $\Delta=0$,
\begin{equation}
    \mu_{144,288}(\Delta=0) = -2(1+\cos(5 \pi/6)) = -0.2679
\end{equation}
As long as $\Delta$ is small relative to $t$ (we find $\Delta/t \sim 0.01$) , this chemical potential does not change very much from the $\Delta=0$ value. Therefore, the two geometries we consider provide an apples-to-apples comparison for $p({\bf k})$. Since $p({\bf k})$ is size invariant, the effective $\Delta$ is size invariant. 

We demonstrate this point more explicitly by fitting to BCS theory at $U=2$,
\begin{equation*}
\begin{split}
    \Delta, \mu = \textrm{argmin}_{\Delta, \mu} \Bigg[ \sum_{\bf k} \frac{ \Big(p({\bf k})  -  p_{\bf k}^{BCS}(\mu,  \Delta) \Big)^2}{\sigma^2_{ p({\bf k})}} + \\ \lambda \Big(n - n^{BCS}(\mu, \delta)\Big)^2  \Bigg]
    \end{split}
\end{equation*}
Here we do an SNR-weighted fit, where $p({\bf k})$ is the sampled average from our neural network wavefunction and $\sigma^2_{ p({\bf k})}$
describes our measurement error for $p({\bf k})$. This is important because $p({\bf k})$ tends to be noisier outside of the Fermi surface. The second term is added to enforce the constraint that the chemical potential $\mu$ agrees with the particle number. We take $\lambda$ to be extremely large. 

\begin{figure*}[t] 
    \centering    \includegraphics[width=0.9\linewidth]{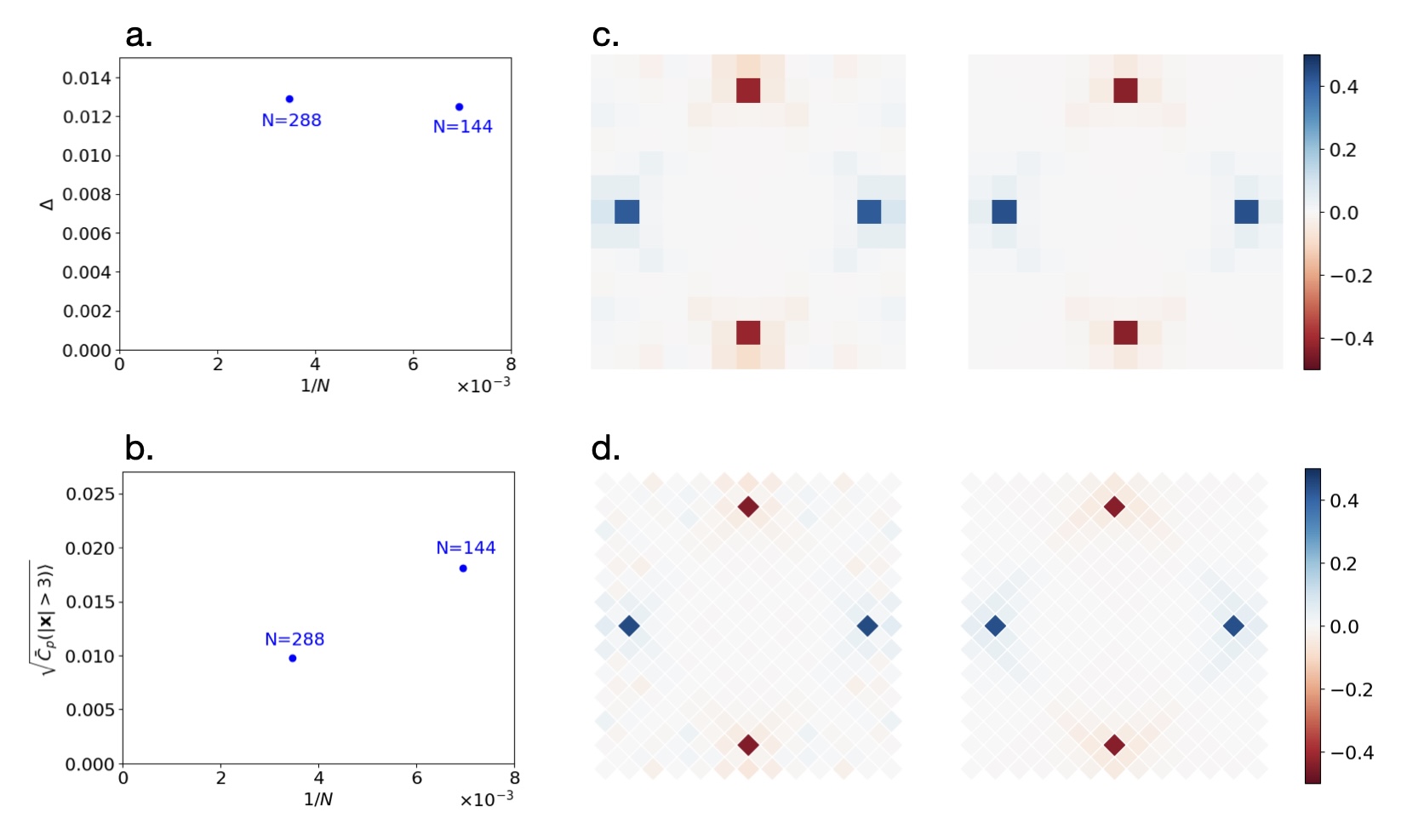}
    \caption{Fitting momentum space pairing to BCS theory. {\bf a.} Computation of $\Delta$ by fitting data to BCS theory on $144$ and $288$ site clusters at $U=2$.  {\bf b.} Estimate of the order parameter from the  
    $\sqrt{{\bar C}_p}$ on 144 and 288 site clusters. {\bf c.} Comparison between numerical data (left) and fit to BCS theory (right) for $U=2$ on 144 site cluster. {\bf d.} Same as {\bf c.} but for 288 site cluster. }
    \label{fig: fittingorder}
\end{figure*}

Fig. \ref{fig: fittingorder}a-b. shows the pairing order parameter, $\Delta$, computed from a fit to BCS theory and ${\bar C}_p(|{\bf x}| > 3)$ for both system sizes at $U=2$. We see that while $\Delta$ is roughly system size invariant, ${\bar C}_p(|{\bf x}| > 3)$ decays strongly with system size. In Fig. \ref{fig: fittingorder}c-d. we show side-by-side color plots of p({\bf k}) computed with NQS, and BCS-theory respectively. We can see that the BCS-theory fit with $\Delta \approx 0.01$ models the data very well. We find that at $U=4$ and $U=6$ this fit unsurprisingly degrades in quality.   
\subsection{How does finite size BCS change with $\Delta$}

Now that we've established that pairing is fairly size invariant based on our numerical data, we document the difficulty of extrapolating a local pairing order parameter. 

\begin{figure*}[htbp] 
    \centering
    \includegraphics[width=0.9\linewidth]{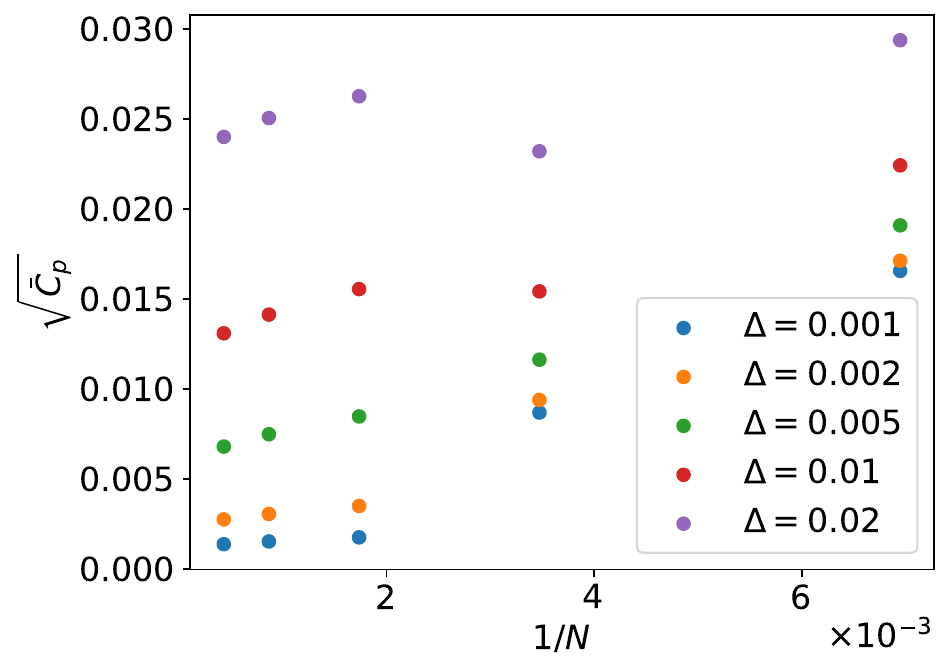}
    \caption{Solving the discrete mean field BCS equations. Here we plot the order parameter estimate from the average pair correlation $\sqrt{{\bar C}_p}$ as a function of system size for different values $\Delta$. Plots are shown for $N=144 \times 2^k$ for $k \in [0,4]$. We are only able to access the two smallest simulation clusters shown.}
    \label{fig: finite size BCS}
\end{figure*}

We solve the mean field BCS equations for fixed $\Delta$ and show how ${\bar C}_p$ scales with system size. First, equation \ref{eqn: number equation} is solved to set the chemical potential, then long range pair correlations are computed from equation \ref{eqn: pair pair BCS} . The results are shown in figure \ref{fig: finite size BCS}

In general, the pair correlations decay with system size, but there are also oscillations. The pair correlations are larger on small systems because the momentum space grid is too coarse to resolve the sharp peak in $\langle p_{\bf k}\rangle$. Therefore, points that are near the Fermi surface are overweighted in the sum in equation \ref{eqn: pair pair BCS}. The occasional upward oscillations occur when a new k-points are introduced that are close to the Fermi-surface. From this data, we see that a system size extrapolation of the traditional d-wave order parameter from ${\bar C}_p$ is a fool's errand, as there is strong finite size effects up to $\sim 1000$ sites. 

\section{Supplementary data}

This section contains additional data that has shaped our understanding of superconductivity in the Hubbard model. 

\subsection{A real space perspective on the BCS-like superconductor} \label{sec: Real space BCS}

\begin{figure*}[t] 
    \centering    \includegraphics[width=0.9\linewidth]{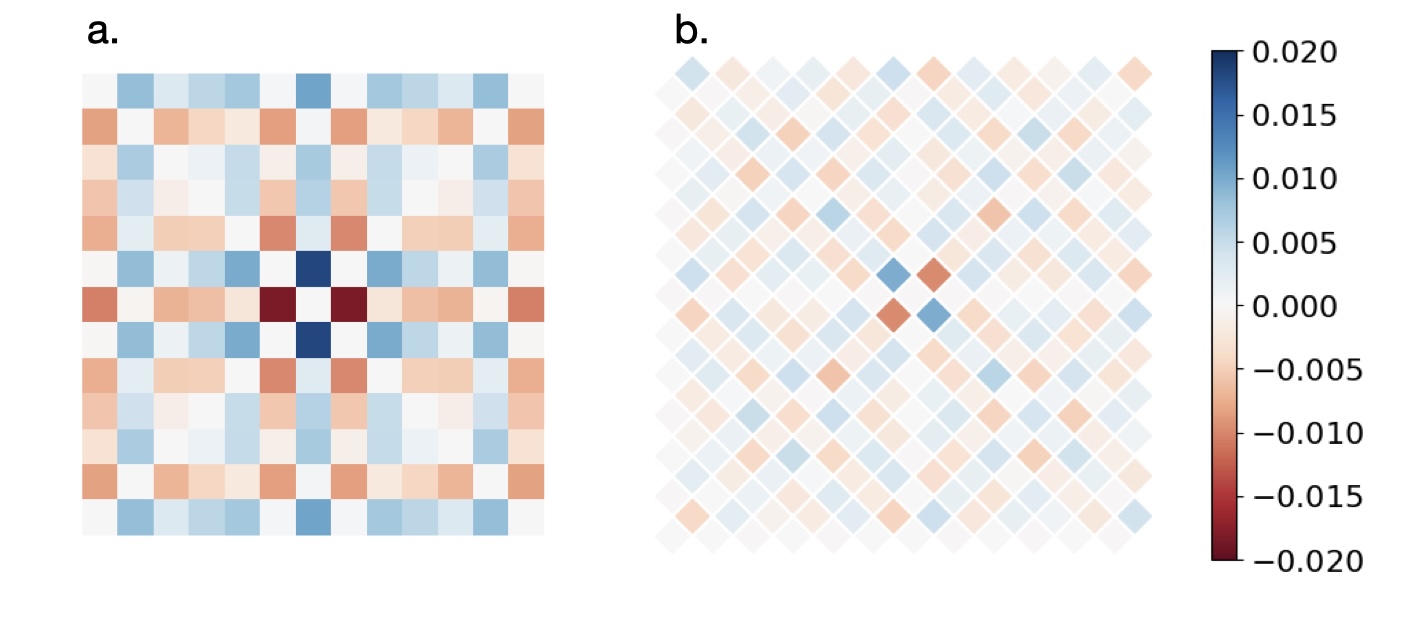}
    \caption{Proxy for real space pairing amplitude $p({\bf x})$ (Eqn. \ref{eqn: real space pairing proxy}) for pair with displacement ${\bf x}$ at $U=4, \delta = 1/6$. {\bf a.} 144 site cluster {\bf b.} 288 site cluster. The nearest neighbors on the $288$ site cluster are rotated by $45 \degree$ relative to the $144$ site cluster}
    \label{fig: real space pairing appendix}
\end{figure*}

In the main text we showed that the momentum space pairing, $p({\bf k})$ is fairly size invariant, while the pair correlations for local d-wave pairs $C_p({\bf x})$ decay with system size. We can understand this effect from a real space perspective by introducing a real space pairing amplitude $p({\bf x})$, which acts as a proxy for the probability of annihilating a pairs with displacement ${\bf x}$. We compute $p({\bf x})$ by Fourier transforming $p({\bf k})$
\begin{equation} \label{eqn: real space pairing proxy}
    p({\bf x}) = \frac{1}{N} \sum_{\bf k} e^{i {\bf k} \cdot {\bf x}} p({\bf k})
\end{equation}
We plot $p({\bf x})$ on $144$ and $288$ site clusters at $U=4$ in Fig. \ref{fig: real space pairing appendix}. On the 288 site cluster the nearest neighbors are rotated by $45 \degree$ relative to the 144 site cluster. Here, we see that the pairs are almost completely de-localized. When the system size is doubled, the pairs delocalize further and the effective amplitude of annihilating a particular pair is halved. This leads to a $1/N^2$ scaling in pair correlations, which create and annihilate a next-neighbor pair.

\subsection{Size invariance across different U} \label{sec: size invariance appendix}

\begin{figure*}[t] 
    \centering \includegraphics[width=0.9\linewidth]{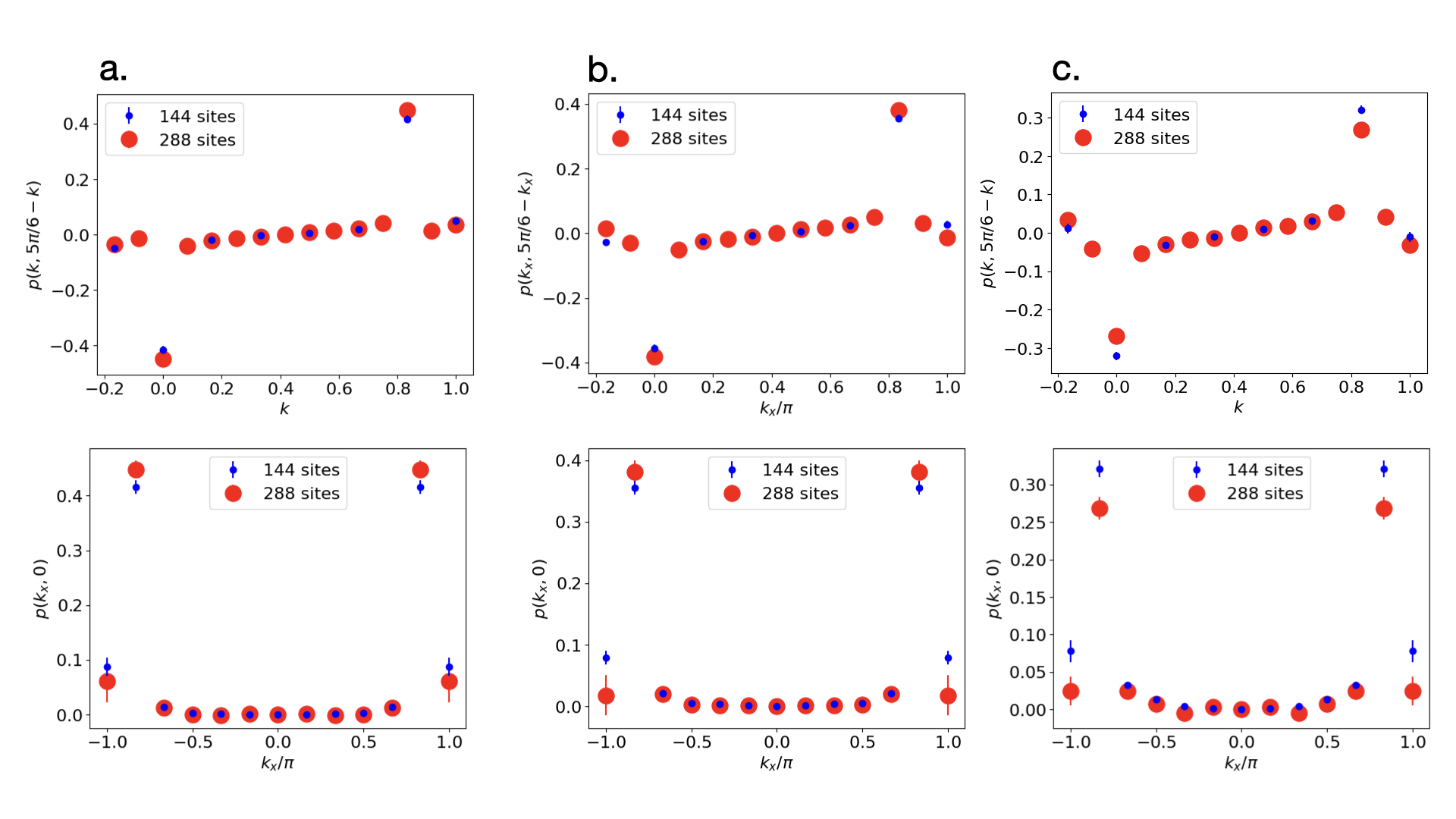}
    \caption{Finite size scaling of $p({\bf k})$ for different U. 1-D cuts of p({\bf k}) along same axes as \ref{fig: pure hubbard}e.-f. for 144 and 288 site clusters at $U=2$ ({\bf a.}), U=4({\bf b.}), and $U=6$({\bf c.})}
    \label{fig: SC size effect $U$}
\end{figure*}

For all values of $U$ that we study, the momentum space pairing, $p({\bf k})$ is roughly system size invariant. In figure \ref{fig: SC size effect $U$}, $p({\bf k})$ is shown on $144$ and $288$ site clusters for different values of $U$. By $U=6$ the peak in $p({\bf k})$ has become much broader, such that the shape can no longer be explained by BCS theory, but the size invariance remains.  

\subsection{Momentum space pairing and finite size effects for the half-filled stripe superconductor} \label{sec: stripy sc appendix}

We show a similar momentum space pairing analysis for the half-filled stripe phase to indicate why the pairing order parameter has likely converged. As shown in \ref{fig: stripemain}{\bf f.} the asymptotic value of the pair correlation functions is roughly invariant going from a 128 to a 256 site cluster. 

\begin{figure*}[htbp] 
    \centering    \includegraphics[width=0.9\linewidth]{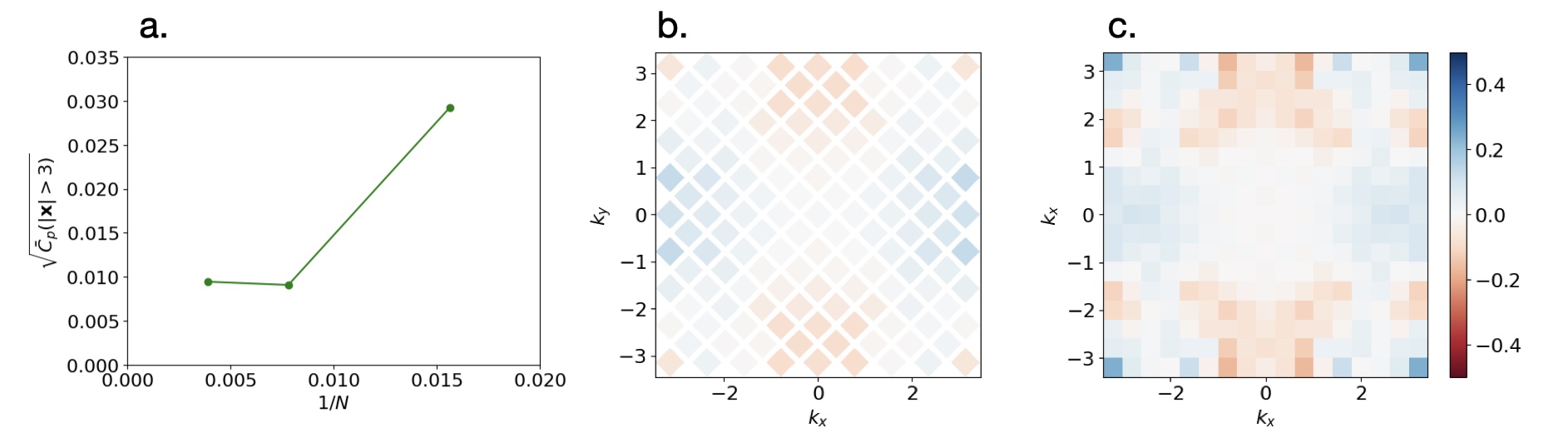}
    \caption{Finite size effects in the half-filled stripe phase. {\bf a.} Estimate of the order parameter from the square root of the asymptotic pair correlation function for  three different system sizes $N=64, 128, 256$. {\bf b.} Momentum space pairing $p({\bf k})$ for the half-filled stripe on a $128$ site cluster {\bf c.}  Momentum space pairing $p({\bf k})$ for the half-filled stripe on a $256$ site cluster.}
    \label{fig: momentumspacestripy}
\end{figure*}

We can also look at $p({\bf k})$ (Eqn. \ref{eqn: momentum space pairing proxy}), and see that it is also fairly system size invariant between the $128$ and $256$ site clusters. In Fig. \ref{fig: momentumspacestripy} we plot the effective order parameter, $\sqrt{\bar{C}_p}$ as a function of three system sizes, as well as $p({\bf k})$ on $128$ and $256$ site clusters.
We see that $p({\bf k})$ has no strong peaks, which indicates that the integral is unlikely to change much as more k-points are introduced.

\subsection{Partially filed stripe normal state}
\begin{figure*}[t] 
    \centering
    \includegraphics[width=0.9\linewidth]{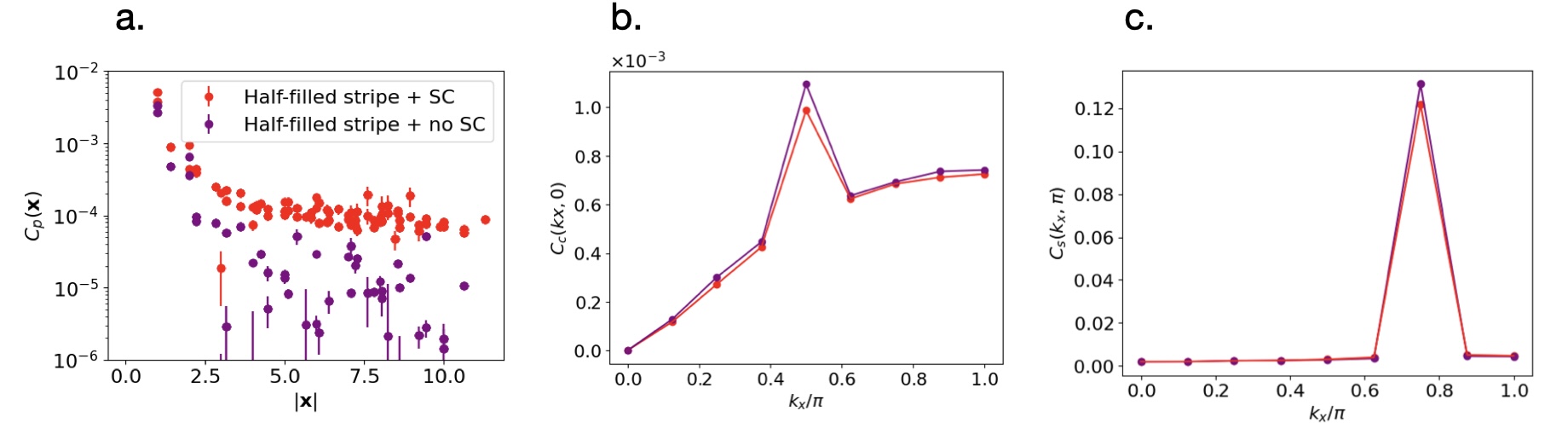}
    \caption{Comparing correlations between the partially-filled stripe (red) and it's normal state (purple) {\bf a.} Pair correlations $C_p({\bf x})$ as a function of displacement $|{\bf x}|$. {\bf b.} Charge structure factor, $C_c({\bf k})$, plotted along the cut $k_x = 0$ {\bf c.} Spin structure factor, $C_s({\bf k})$, plotted along the cut $k_x = \pi$}
    \label{fig: stripynormal}
\end{figure*}

By starting our optimization from an unpaired trial wavefunction, we can also find a ``normal state" for the partially-filled stripe phase. This phase has $\sim 1 \%$ higher energy than the superconducting phase, and very similar charge/spin correlation functions. In figure \ref{fig: stripynormal}a., we plot the pair correlations for both phases, showing that the normal state has no superconductivity. In figure \ref{fig: stripynormal}b.-c., we plot relevant 1-D cuts of the charge and spin structure factors. We see that the normal state has slightly enhanced charge/spin structure factor peaks.   
\subsection{Effects of changing lattice geometry} \label{sec: geometry}

In order to understand the effect of lattice geometry, we also trained models for the superconducting and stripe phases on a $24 \times 4$ torus. We find that the energies are extremely comparable, with the stripe phase slightly favored,
\begin{table}[H]
    \centering
    \begin{tabular}{|c|c|c|} \hline
         Phase & {\bf Filled Stripe} &  Superconductor \\ \hline
        Energy & {\bf -1.07872(3)}&-1.07779(5) \\ \hline \end{tabular}
    \caption{Energies per site of competing phases at $U=4$ on a $24 \times 4$ torus. The lower energy phase is bolded.}
    \label{tab:energies asymmetric}
\end{table}

This is a notably different result from the $12 \times 12$ torus (table \ref{tab:energy small}) where the superconductor is $\sim 0.5 \%$ lower in energy. Therefore, we see that boundary effects can profoundly affect the location of the stripe transition. While this transition is between $U=6$ and $U=10$ on the $12 \times 12$ cluster (see Sec. \ref{sec: changing U}), it is below $U=4$ on the $24 \times 4$ torus.  

In figure \ref{fig: pair asymmetric} we show pair correlations for the superconducting phase on this geometry. We see that the pair correlation function oscillates, precluding an obvious way to estimate long range order. These differences have motivated us to focus our simulations on symmetric tori. 

\section{Optimization details}

\subsection{Training procedure} \label{sec: training procedure}

We optimize our HFPS wavefunction in a two step procedure as done in reference \cite{chen2025pfaffian}. First we optimize the mean field using exact gradient techniques. Using the mean field orbitals $F$, the two-body expectation values $\langle c^\dagger_{\bf x, \sigma} c_{\bf x', \sigma'} \rangle $, $\langle c_{\bf x, \sigma} c_{\bf x', \sigma'} \rangle $ can be computed. Expectation values of four body operators can then be computed with Wick's theorem. This allows us to compute an energy function $E(F,H)$, where $H$ is defined in Eqn. \ref{eqn: mean field ham}, and use automatic differentiation to update F in order to move downhill in the energy landscape, 
\begin{equation}
\Delta F = -\frac{d}{dF} ( E(F,H))
\end{equation}
We typically minimize this function for 50K steps until the energy is fairly converged. 

Typically this optimization is much more efficient when $F$ is in a block circulant form corresponding to a particular mean field order.  For the t-t' model at $\delta = 1/8$, we try both fully-filled and half-filled stripes. For higher dopings, we always use a unit cell that accommodates a fully filled stripe. 

It is important to add a d-wave pinning field to $H$ in order to represent a superconducting phase. This is because the pairing only enters $E(F,H)$ through a quadratic form.
\begin{equation*}
\begin{split}
   U_0 \langle c^\dagger_{\bf x \uparrow} c_{\bf x \uparrow} c^\dagger_{\bf x \downarrow} c_{\bf x \downarrow} \rangle = \\ U_0 \langle c^\dagger_{\bf x \uparrow} c^\dagger_{\bf x \downarrow} \rangle \langle c_{\bf x \downarrow}  c_{\bf x \uparrow}\rangle + ... = U_0 \langle c_{\bf x \downarrow}  c_{\bf x \uparrow}\rangle^2 + ...
   \end{split}
\end{equation*}
For $U_0 > 0$ this is always minimized when $\langle c_{\bf x \downarrow}  c_{\bf x \uparrow}\rangle = 0$. 

\subsection{Hyperparameters}

All neural network were trained via imaginary time evolution starting from the mean field states specified in the main text. We implement this  using the MinSR algorithm \cite{Chen_NP24_MinSR,Rende_CP24_SRt}, with $r_{tol} = 10^{-12}$. The learning rate for all calculations was $\eta = 0.005$. For each calculation we used $8$ hidden fermions, $18$ features and $8$ residual layers of HFPS + GCNN architecture described in reference \cite{chen2025pfaffian}.

\begin{figure*}[htbp]
    \centering
    \includegraphics[width=0.9\linewidth]{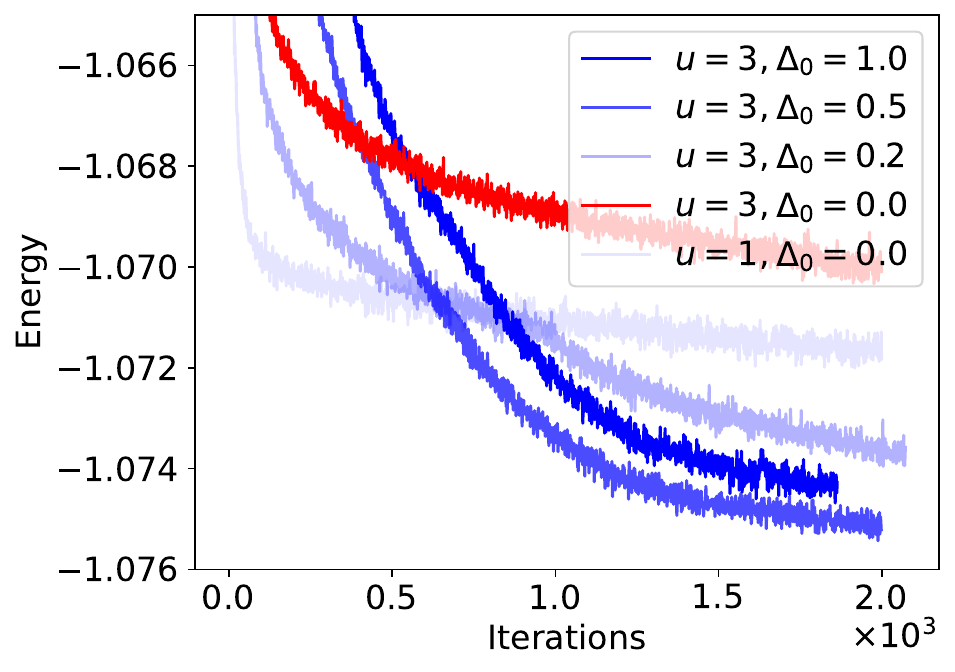}
    \caption{Training curves of simulations for wavefunctions initialized with a mean field trained on different Hamiltonians for $U=4$, $\delta = 1/6$ on a $144$ site cluster. The $U_0=3$, $\Delta_0 = 0$ initialization converges to a stripe phase, while all of the other initialization converge to superconducting states}
     \label{fig: energies}
\end{figure*}
 
\begin{figure*}[htbp]
    \centering
    \includegraphics[width=0.9\linewidth]{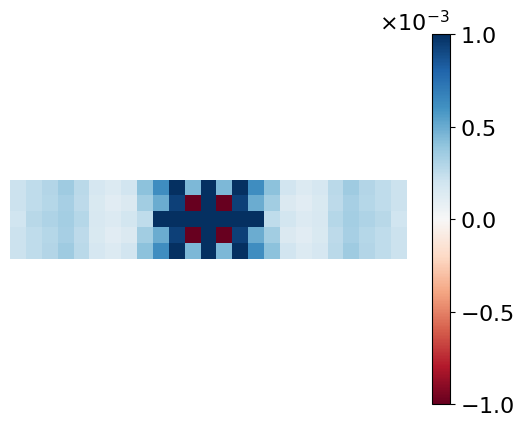}
    \caption{D-wave pair correlations for the superconducting phase trained on a $24 \times 4$ torus at $U=4$, $\delta = 1/6$}
     \label{fig: pair asymmetric}
\end{figure*}

\clearpage

\begin{table*}[ht!]
    \centering
    \begin{tabular}{|c|c|c|} \hline
      System & Training steps & Optimization time \\ \hline 
      $16 \times 16$, $\delta = 1/8$, filled & 2000 & 5000 \\ \hline
      $16 \times 16$, $\delta = 1/8$, half-filled & 2000 & 5000 \\ \hline
      $12 \times 12$, $\delta =  1/6$ & 1000 & 500 \\ \hline
      $12 \sqrt{2} \times 12 \sqrt{2}$, $\delta =  1/6$ & 2500 & 10000 \\ \hline
      $10 \times 10$, $\delta =  1/5$ & 1000 & 300 \\ \hline
      $10 \sqrt{2} \times 10 \sqrt{2}$ & 2500 & 5000 \\ \hline    
    \end{tabular}
    \caption{Rough estimates of training times in NVIDIA H200 gpu hours.}
    \label{tab: half filling}
\end{table*}

\begin{table*}[]
    \centering
    \begin{tabular}{|c|c|c|c|} \hline
         Phase & Filled Stripe & Half-filled stripe + No SC &  Half-filled stripe + SC \\ \hline
        $t^\prime=-0.3$ & N/A & -0.7245(3)  & -0.7375(1)   \\ \hline
        $t^\prime=-0.2$ & -0.7261(1) & N/A & -0.7360(1)  \\ \hline
        $t^\prime=-0.1$ & -0.7367(1) & N/A & -0.7370(1)  \\ \hline
        $t^\prime=0.0$ & -0.7515(1) & -0.7271(3) & -0.7401(1)  \\ \hline
        \end{tabular}
    \caption{Energy per site of competing phases on a $256$ site for various values of $t^\prime$ shown in Fig \ref{fig: stripemain}}
    \label{tab:energies stripe}
\end{table*}

\begin{table*}[]
    \centering
    \begin{tabular}{|c|c|c|c|} \hline
         Phase & Filled Stripe & Half-filled stripe + No SC &  Half-filled stripe + SC \\ \hline
        $t^\prime=-0.3$ & N/A & 0.1487 & 0.07968  \\ \hline
        $t^\prime=-0.2$ & 0.05810 & N/A & 0.07623  \\ \hline
        $t^\prime=-0.1$ & 0.05291 & N/A & 0.07245  \\ \hline
        $t^\prime=0.0$ & 0.06379 & 0.1450 & 0.07838  \\ \hline
        \end{tabular}
    \caption{Variance per site of competing phases on a $256$ site for various values of $t^\prime$ shown in figure \ref{fig: stripemain}{\bf c.}}
    \label{tab:variance stripe}
\end{table*}

\begin{table*}[]
    \centering
    \begin{tabular}{|c|c|c|} \hline
         $U$ & E/N & $\sigma^2/N$ \\ \hline
        2 & -1.29500(3) & 0.02243 \\ \hline
        4 & -1.07508(4) & 0.04779  \\ \hline
        6 & -0.92516(5) & 0.07778 \\ \hline
        10 & -0.76179(5) & 0.06394 \\ \hline
        15 & -0.66690(5) & 0.05883 \\ \hline       
        \end{tabular}
    \caption{Energy and variance per site of the lowest energy wavefunction on the $144$ site cluster at various $U$}
    \label{tab:energy small}
\end{table*}

\begin{table*}[]
    \centering
    \begin{tabular}{|c|c|c|} \hline
        $U$  & Energy & Variance \\ \hline
        2 & -1.29563(3) & 0.04165 \\ \hline
        4 & -1.07256(5) & 0.11256  \\ \hline
        6 & -0.91859(6) & 0.19935 \\ \hline
        \end{tabular}
    \caption{Energy and variance per site of the lowest energy wavefunction on $288$ site cluster at various $U$}
    \label{tab:energies big}
\end{table*}

\end{document}